\documentclass[aps,floatfix,prl,superscriptaddress,twocolumn]{revtex4-1}
\usepackage{bbm}
\usepackage{graphicx}
\usepackage{dcolumn}
\usepackage{subfigure}
\usepackage{amsmath}
\usepackage{amsfonts}
\usepackage{appendix}
\usepackage{feynmf}
\usepackage{hyperref}
\usepackage{eepic}
\usepackage{lipsum}
\usepackage{enumerate}
\usepackage{amssymb}
\usepackage[english]{babel}

\usepackage{attachfile}
\newcommand{\ua}{\uparrow}
\newcommand{\da}{\downarrow}
\newcommand{\ra}{\rightarrow}

\newcommand{\bk}{\mathbf k}
\newcommand{\bR}{\mathbf R}

\newcommand{\bq}{\mathbf q}
\newcommand{\br}{\mathbf r}

\usepackage{times}

\begin{document}
\title{Anomalous Hall effect in single-band chiral superconductors from impurity superlattices}
\author{Yu Li}
\address{Shenzhen Institute for Quantum Science and Engineering \& Guangdong Provincial Key Laboratory of Quantum Science and Engineering, Southern University of Science and Technology, Shenzhen 518055, Guangdong, China}
\affiliation{Kavli Institute for Theoretical Sciences, University of Chinese Academy of Sciences, Beijing 100190, China}
\author{Zhiqiang Wang}
\affiliation{Department of Physics and Astronomy, McMaster University, Hamilton, Ontario L8S 4M1, Canada}
\affiliation{James Franck Institute, University of Chicago, Chicago, Illinois 60637, USA}
\author{Wen Huang}
\email[]{huangw3@sustech.edu.cn}
\address{Shenzhen Institute for Quantum Science and Engineering \& Guangdong Provincial Key Laboratory of Quantum Science and Engineering, Southern University of Science and Technology, Shenzhen 518055, Guangdong, China}
\date{\today}

\begin{abstract}
Unlike anomalous quantum Hall insulators, clean single-band chiral superconductors do not exhibit intrinsic Hall effect at the one-loop approximation. Finite ac Hall conductance was found to emerge beyond one-loop, such as with vertex corrections associated with extrinsic random impurity scatterings. In this paper, we investigate the effect of impurities embedded in single-band chiral superconductors in a superlattice pattern, instead of in random distributions. The impurity-induced Bogoliubov quasiparticle bound states hybridize to form subgap bands, constituting an emergent low-energy effective theory whose Hall effect can be studied with ease. We demonstrate that the occurrence of the Hall effect depends on the superlattice geometry and on the parity of the chiral pairing. In particular, due to the mixed particle-hole character of the subgap states, the Hall conductance may arise at the one-loop level of the current-current correlator in our effective model. Our theory provides a new insight into the impurity-induced anomalous Hall effect in chiral superconductors.
\end{abstract}

\maketitle

{\it Introduction}-- Topological chiral superconductors are classified by a topological invariant -- the Chern number, and they exhibit protected chiral edge modes. Odd-parity chiral superconductors (e.g. chiral p-, f-wave, etc) may further support half-quantum vortices that host Majorana zero modes~\cite{Volovik:99,Read:00}. These excitations obey non-Abelian braiding statistics and could therefore be utilized for topological quantum computation~\cite{Ivanov:01,Kitaev:03,Nayak:08}.

The time-reversal symmetry breaking of the chiral pairings can be detected in polar Kerr effect measurements, where a linearly polarized light normally incident on the superconductor is reflected with a rotated polarization. Signatures of Kerr rotation have been reported in several unconventional superconductors, including Sr$_2$RuO$_4$~\cite{Xia:06}, UPt$_3$~\cite{Schemm:14} and URu$_2$Si$_2$~\cite{Schemm:15}. Such an effect is closely related to the anomalous quantum Hall effect. However, unlike in an anomalous Hall insulator, the effect is not expected in a clean and uniform single-band chiral superconductor~\cite{Read:00,Lutchyn:09}. This could be understood in the following simple terms. The pairing potential $\Delta_{\bk}$, whose $\bk$-dependence describes the relative motion between the paired electrons, does not generate center-of-mass motion for the Cooper pair. Thus the current operator of a superconductor contains no contribution originating from $\Delta_{\bk}$. Consequently, the Hall conductance is not directly related to the Berry curvature of the Bogoliubov quasiparticles and it in fact vanishes at the one-loop approximation. 

Nonetheless, vertex corrections, such as those arising from extrinsic impurity scatterings \cite{Sinitsyn:08,Goryo:08,Goryo:10,Lutchyn:09,Li:15,Konig:17} and certain intrinsic superconducting collective modes \cite{Yip:92,Roy:08,Lutchyn:08}, have both been shown to induce finite Hall conductance. Separate intrinsic mechanisms exist for multiband superconductors, but those involve interband Cooper pairing~\cite{Taylor:12,Annett:12,Wang:17,Brydon:19,Komendova:17}. Thus far, whether these effects could quantitatively explain the observed Kerr rotation is still debated \cite{Kallin:16,Wysokinski:19}.

In previous studies, the leading order impurity effects are captured by the so-called skew-scattering diagrams \cite{Goryo:08,Goryo:10,Lutchyn:09,Konig:17}. However, these studies only capture the continuum state contribution, and some important microscopic details are absent in the diagrammatic treatment. In particular, individual impurities are known to induce subgap quasiparticle bound states~\cite{Okuno:99,Balatsky:06}. How such low-energy states influence the electromagnetic response of the system remains largely unexplored and is the focus of the present study. In so doing, we uncover a new perspective on the impurity-induced Hall effect in chiral superconductors. 

\begin{figure}
\includegraphics[width=0.43\textwidth]{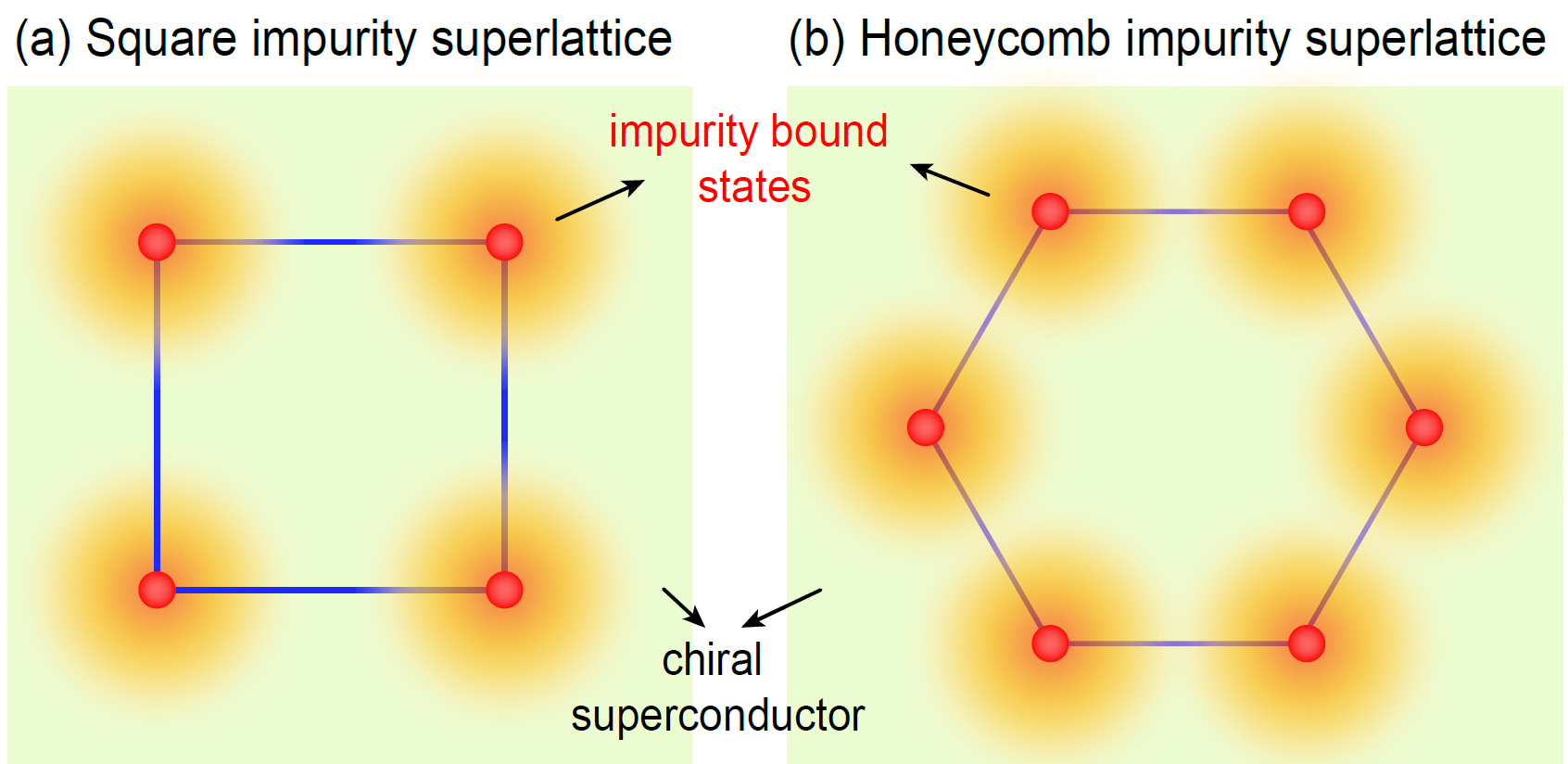}
\caption{(a) square and (b) honeycomb impurity superlattices embedded in a chiral superconductor.}
\label{fig1}
\end{figure}

To facilitate our discussions, we imagine depositing impurities on the underlying chiral superconductors in a superlattice pattern. Due to the chiral nature of the pairing, the bound states from different impurity sites hybridize in a peculiar fashion that depends on their relative position. We construct a low-energy effective theory of the emergent subgap bands on the superlattice and study the resultant Hall response. Despite having similar appearance, the new effective Hamiltonian differs from the original BdG Hamiltonian in a fundamental way, that the components of the new spinor basis are no longer purely electron or hole, but rather a linear superposition of both. This mixed particle-hole character has profound consequences on transport properties, and we shall demonstrate finite Hall conductance at the one-loop level.

We consider several representative impurity superlattice geometries, and show that the resultant physics is model-dependent. For example, while the Hall conductance is generically nonvanishing on a honeycomb superlattice in any underlying chiral pairing, it vanishes for square and triangular superlattices embedded in even-parity chiral superconductors, such as $d$- and $g$-wave states.

{\it Impurity states and impurity superlattice}-- In the Nambu spinor basis $\hat{\varphi}(\br)=(c_{\br\ua},c^{\dagger}_{\br\da})^\mathsf{T}$, the underlying single-band chiral superconducting state is described by the continuum Bogoliubov-de Gennes (BdG) Hamiltonian $H^{(\text{bulk})}_{\text{BdG}}=\int d\br d\br^{\prime} \hat{\varphi}^{\dagger}(\br)\hat{H}^{(\text{bulk})}_{\text{BdG}}(\br,\br^{\prime})\hat{\varphi}(\br^{\prime})+\text{H.c.}$, in which
\begin{equation}
\hat{H}^{(\text{bulk})}_{\text{BdG}}(\br,\br^{\prime})=\left(\begin{array}
[c]{cc}
\delta_{\br,\br^{\prime}}(-\frac{\nabla^2_{\br^{\prime}}}{2m_e}-\mu) & \Delta(\br,\br^{\prime}) \\
\Delta^{*}(\br,\br^{\prime}) & \delta_{\br,\br^{\prime}}(\frac{\nabla^2_{\br^{\prime}}}{2m_e}+\mu)
\end{array}\right)
\label{eqn1}
\end{equation}
where $c_{\br}~ (c^{\dagger}_{\br})$ stands for the electron annihilation (creation) operators, $m_e$ and $\mu$ are the electron mass and the chemical potential, respectively. The off-diagonal term $\Delta(\br,\br^{\prime})=g(|\br-\br^{\prime}|)e^{il\theta_{\br-\br^{\prime}}}$ is the chiral pairing potential, where $\theta_{\br}$ is the azimuthal angle of $\br$, and $g(|\br-\br^{\prime}|)$, assumed to be a certain (unimportant) decaying function of $|\br-\br^{\prime}|$, describes the spatial profile of the Cooper pair wavefunction. Here $l$ denotes the order of the chiral pairing, i.e., the Cooper pair angular momentum, which takes the values $1,2,\cdots$ for $p_x+ip_y$, $d_{x^2-y^2}+id_{xy}$ pairings, etc. Notice that we have assumed a uniform order parameter independent of the Cooper pair center-of-mass position, $(\br+\br^{\prime})/2$. Consideration of spatial variations around impurities does not qualitatively alter our conclusion.

Impurities in chiral superconductors are known to induce bound states. Consider first a single-impurity at $\mathbf{R}=0$, described by a delta-function-like potential $U\delta(\mathbf{r-R})\tau_3$ where $U$ is the impurity strength and $\tau_3$ is the third component of the Pauli matrices operating in the Nambu space. The bound state wavefunctions take the forms $\psi_{+}(\mathbf{r})=(u(\mathbf{r}),\upsilon(\mathbf{r}))^\mathsf{T}=(u_r,e^{-il\theta_{\mathbf{r}}}\upsilon_r)^\mathsf{T}$ and $\psi_{-}(\mathbf{r})=(-\upsilon^*(\mathbf{r}), u^*(\mathbf{r}))^\mathsf{T}$~\cite{supp}. Here the `$+$' and `$-$' designate, respectively, the state with subgap energy $+E_0$ and the other with $-E_0$, where $E_0<\Delta_0$ and $\Delta_0$ denotes the superconducting gap. These two states are related by particle-hole symmetry, but the detailed forms of $u_r$ and $v_r$ are model-dependent and are not constrained by any other symmetry, except that they shall in general decay as $e^{-r/\xi}/\sqrt{k_\text{F}r}$ sufficiently far away from the impurity center. Here $k_\text{F}$ is the Fermi momentum and $\xi$ the superconducting healing length. In the following, we shall assume a sizable impurity strength such that $E_0\ll \Delta_0$~\cite{Kaladzhyan:16}, under which circumstance the low-energy theory associated with these bound states are well separated from the continuum spectrum.

\begin{figure}
\includegraphics[width=0.35\textwidth]{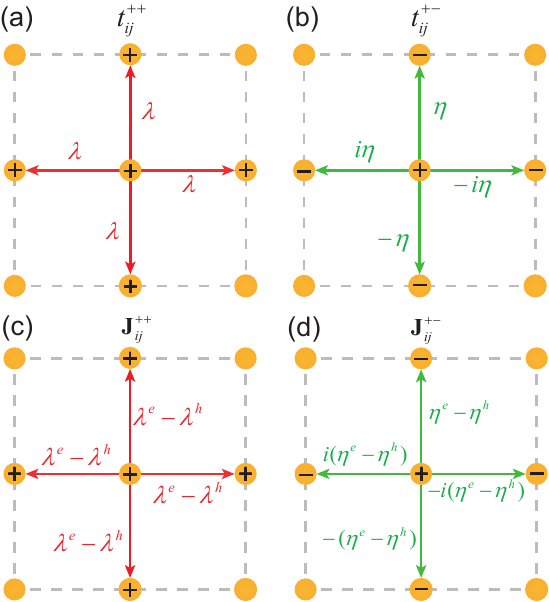}
\caption{(a) (b) Tight-binding construction of a square impurity superlattice immersed in a chiral p-wave superconductor. Note the relation $\lambda=\lambda^\Delta+ \lambda^e+\lambda^h$ and $\eta=\eta^\Delta+ \eta^e+\eta^h$. (c) (d) The current operator on the superlattice. The `$+$' and `$-$' symbols on the sites label the impurity bound states, and arrows indicate the reference direction of hopping or current flow.}
\label{fig:TightBinding}
\end{figure}

On an impurity lattice where the interlattice spacing $R_0$ is larger than $\xi$, the above-stated bound state wavefunctions on each single site still constitute a good approximation. States from neighboring impurity sites `hybridize' via the microscopic kinetic hopping and pairing in the original Hamiltonian Eq.~(\ref{eqn1}). Written in the second quantized form where $c^\dagger_{i,\pm}$ ($c_{i,\pm}$) denote the creation (annihilation) of the respective bound states on each site, an emergent low-energy tight-binding model on the superlattice reads $H^{\text{eff}}=\sum_{i,j}\hat{\Psi}^{\dagger}_{i}\left[E_0\delta_{ij}\sigma_3+ \hat{h}_{ij}(1-\delta_{ij})\right]\hat{\Psi}_{j}+\text{H.c.}$, with the Pauli $\sigma$-matrices operating in the space spanned by $\hat{\Psi}_{i}=(c_{i,+},c_{i,-})^{\mathsf{T}}$, and
\begin{equation}
\hat{h}_{ij}=\left(
\begin{array}
[c]{cc}%
t^{++}_{ij} & t^{+-}_{ij}\\
t_{ij}^{-+} & t^{--}_{ij}%
\label{eq:Heff}
\end{array}
\right),\\
\end{equation}
in which
\begin{equation}
t^{\mu\nu}_{ij}=\int d\br d\br^{\prime}\psi^{\dagger}_{\mu}(\mathbf{r-R}_i) \hat{H}_{\text{BdG}}(\br,\br^{\prime})\psi_{\nu}(\mathbf{r^{\prime}-R}_j).
\label{eq:t12}
\end{equation}
It is obvious that the hopping of the bound states could arise from both the kinetic and pairing terms in the underlying microscopic Hamiltonian. A detailed analysis of the hopping integrals can be found in the Supplementary~\cite{supp}, which we summarize below and in Fig.~\ref{fig:TightBinding} (a) and (b). The hybridization between the `$+$' (`$-$') states satisfy $t^{++}_{ij}=-t^{--}_{ij}=\lambda_{ij}$, where $\lambda_{ij}$ is a real constant determined by the separation $|\mathbf{R}_j-\mathbf{R}_i|$.
On the other hand, the integral between `$+$' and `$-$' states has the relation $t^{+-}_{ij}=(t^{-+}_{ij})^\ast=\eta_{ij}$, where $\eta_{ij} = |\eta_{ij}|e^{il\theta_{\mathbf{R}_j-\mathbf{R}_i}}$~\cite{footnote2}. It thus depends on both the relative position between the two sites and the order of the chiral pairing.

Our later analyses of the current operators require distinguishing in Eq.~(\ref{eq:t12}) contributions originating from the pairing, the electron and hole kinetic hopping processes, i.e. $\lambda_{ij}=\lambda^\Delta_{ij}+\lambda^e_{ij}+\lambda^h_{ij}$ and $\eta_{ij} = \eta^\Delta_{ij}+\eta^e_{ij}+\eta^h_{ij}$. The kinetic part of $\eta_{ij}$, $\eta^e_{ij}+\eta^h_{ij}$, deserves special attention. Written explicitly,
\begin{widetext}
\begin{eqnarray}
\eta_{ij}^{e}+\eta_{ij}^{h}   &=&\int d\mathbf{r}d\mathbf{r}^{\prime}\left[
-u_{\left\vert \mathbf{r-R}_{i}\right\vert } \delta_{\mathbf{r,r}^{\prime}}\left(  -\nabla_{\mathbf{r}^{\prime}}^{2}/2m_{e}-\mu\right)%
  e^{il\theta_{\mathbf{r}^{\prime}\mathbf{-R}_{j}}}%
\upsilon_{\left\vert \mathbf{r}^{\prime}\mathbf{-R}_{j}\right\vert }  +e^{il\theta_{\mathbf{r-R}_{i}}}\upsilon_{\left\vert \mathbf{r-R}%
_{i}\right\vert }  \delta_{\mathbf{r}^{\prime}\mathbf{,r}}\left(  \nabla_{\mathbf{r}^{\prime}}^{2}/2m_{e}+\mu\right)
u_{\left\vert \mathbf{r}^{\prime}\mathbf{-R}_{j}\right\vert }\right]   \nonumber\\
 &=&\left(  1+e^{il\pi}\right)  \int d\mathbf{\tilde{r}}d\mathbf{\tilde{r}}^{\prime}%
u_{\tilde{r}} \delta_{\mathbf{\tilde{r},\tilde{r}}^{\prime}-(\mathbf{R}_{i}-\mathbf{R}_{j}%
)}\left(  \nabla_{\mathbf{\tilde{r}}^{\prime}}^{2}/2m_{e}%
+\mu\right)  e^{il\theta_{\mathbf{\tilde{r}}^{\prime}}}\upsilon_{\tilde{r}^\prime},
\label{eq:tPMeh}
\end{eqnarray}
\end{widetext}
where we have performed partial integration and substitution of variables to obtain the second line, and the two terms in $1+e^{il\pi}$ are associated with $\eta^e_{ij}$ and $\eta^h_{ij}$, respectively. The relation $|\eta^e_{ij}|=|\eta^h_{ij}|$ is a consequence of the particle-hole symmetry between the `$+$' and `$-$' states. For $l$ odd, $\eta^e_{ij} = - \eta^h_{ij}$, hence the kinetic contribution vanishes if the underlying chiral pairing has odd-parity; for $l$ even, by contrast, $\eta^e_{ij} = \eta^h_{ij}$. We shall later see that the corresponding current operators have the opposite even and odd $l$-dependence. Finally, it is easy to check that the relation $\eta_{ij} = |\eta_{ij}|e^{il\theta_{\mathbf{R}_j-\mathbf{R}_i}}$ also holds for the individual constituents of $\eta_{ij}$.

As a concrete example, in a chiral $p$-wave superconductor, a square impurity superlattice with up to nearest-neighbor hopping has the following effective Hamiltonian,
\begin{equation}
\hat{H}^\text{eff}_{\bk}= \mathcal{E}_{3\bk} \sigma_3 + \mathcal{E}_{1\bk} \sigma_1 -\mathcal{E}_{2\bk} \sigma_2
\label{eq:PHeffk}
\end{equation}
where we have set $R_0=1$ for brevity, $\mathcal{E}_{3\bk} = 2\lambda (\cos k_x + \cos k_y) +E_0$, $\mathcal{E}_{1\bk} =2\eta \sin k_x $ and $\mathcal{E}_{2\bk} =2\eta \sin k_y $. Here, $\lambda$ denotes the nearest-neighbor hopping integrals of $\lambda_{ij}$, and $\eta$ the corresponding counterpart of $|\eta_{ij}|$. Notice the implicit decomposition such as $\eta = \eta^\Delta + \eta^e + \eta^h$ (although $\eta^e + \eta^h=0$ for chiral p-wave). Due to the angle dependence of the complex off-diagonal hopping $\eta_{ij}$, Eq.~(\ref{eq:PHeffk}) resembles the form of the underlying chiral p-wave Hamiltonian. The band topology could be engineered by controlling parameters such as the impurity potential and the superlattice constant~\cite{Kaladzhyan:16,Kimme:16}. These hold for higher order chiral superconductors, although further neighbor hybridizations must be considered to make the band topology transparent. In like manner, impurity chains immersed in odd-parity chiral states support an emergent 1D p-wave model and may give rise to isolated Majorana zero modes at the ends of the chains.

{\it Current operators}-- The mixed particle-hole nature of each of the spinor component in $\hat{\Psi}$ (i.e. each impurity bound state) has a profound consequence on the particle current operators. Foremost, the portion of the hopping integrals originating from the underlying Cooper pairing, i.e. $\lambda^\Delta$ and $\eta^\Delta$, shall have no contribution, as in the case of clean superconductors. The only contribution stems from the mutually `canceling' electron hopping ($\lambda^e$ and $\eta^e$) and hole hopping ($\lambda^h$ and $\eta^h$). Understandably, if the `$+$' state is purely electron-like and the `$-$' state purely hole-like, $\eta^e=\eta^h=0$, and the resultant current operators resemble those of a clean superconductor.

For the model given in (\ref{eq:PHeffk}), the current operators $\mathbf{J}_{ij}^{++}$ and $\mathbf{J}_{ij}^{+-}$ defined on the superlattice bonds are sketched in Fig.~\ref{fig:TightBinding} (c) and (d). The properties of the $t_{ij}^{\mu\nu}$'s imply the following general relation: $\mathbf{J}^{++}_{ij}=-(\mathbf{J}^{--}_{ij})^\ast$ and $\mathbf{J}^{+-}_{ij}=(\mathbf{J}^{-+}_{ij})^\ast$. Specific to the model in (\ref{eq:PHeffk}), the $x$-component of the current operator reads,
\begin{equation}
\hat{J}^\text{eff}_{x\bk} = \mathcal{J}_{3x\bk} \sigma_3+ \mathcal{J}_{1x\bk} \sigma_1 + \mathcal{J}_{2x\bk} \sigma_2 \,,
\label{eq:Jx}
\end{equation}
where $\mathcal{J}_{3x\bk}=-2(\lambda^e - \lambda^h)\sin k_x$, $\mathcal{J}_{1x\bk}=2(\eta^e - \eta^h)\cos k_x $, and $\mathcal{J}_{2x\bk}=0$. Note that $\mathcal{J}_{2x}$ could be nonzero if further neighbor hoppings are considered. The $y$-component follows similarly and can be found in the Supplementary~\cite{supp}. The cancellation between the electron and hole contributions is evident in these expressions. Notably, although $\eta^e+\eta^h=0$ for odd-parity pairing, the corresponding kinetic contribution to the particle current is finite and scales as $\eta^e-\eta^h=2\eta^e$, such as in $\mathcal{J}_{1x\bk}$. In the case of underlying even-parity pairing, however, since $\eta^e=\eta^h$, $\mathbf{J}^{+-}_{ij} \propto \eta^e-\eta^h=0$ -- suggesting a perfect cancellation between the electron and hole transport. Hence $\mathcal{J}_{1x(y)}$ and $\mathcal{J}_{2x(y)}$ must both vanish in this case.

{\it Anomalous Hall conductivity}-- We are now in position to study the anomalous Hall conductance of our low-energy theory. Within linear response theory, it is given by the antisymmetric part of the $\hat{J}_x-\hat{J}_y$ correlation function $\pi_{xy}(\bq,\omega)$,
\begin{equation}
\sigma_{\text{H}}(\omega)=\frac{i}{2\omega}\lim_{\bq\rightarrow 0}\left[\pi_{xy}(\bq,\omega) -\pi_{yx}(\bq,\omega)\right],
\label{eqn4}
\end{equation}
where, at the one-loop approximation,
\begin{equation}
\begin{aligned}
\pi_{xy}(\bq=0,i\nu_m)=&T\sum_{\bk,i\omega_n}\text{Tr}\left[\hat{J}^\text{eff}_{x\bk} \hat{G}(\bk,i\omega_n+i\nu_m)\right.\\
& \left. \times\hat{J}^\text{eff}_{y\bk} \hat{G}(\bk,i\omega_n) \right],
\end{aligned}
\label{eqn5}
\end{equation}
where $T$ is the temperature, $\omega_n=(2n+1)\pi T$ and $\nu_m=2m\pi T$ are, respectively, the fermionic and bosonic Matsubara frequencies, and $\hat{G}(\bk,i\omega_n)=(i\omega_n-\hat{H}^{\text{eff}}_{\bk})^{-1}$ stands for the impurity-band Green's function. For the square lattice model introduced above, we arrive at the following,

\begin{equation}
\sigma_{\text{H}}(\omega+i\delta)=\sum_{\mathbf{k}}\frac{f_{\bk} }{E_{\mathbf{k}}\left[  (\omega+i\delta)^{2}-4E_{\mathbf{k}}^{2}\right]  } \,,
\label{eq:HallFinal}
\end{equation}
where $E_{\mathbf{k}}=\sqrt{\mathcal{E}^2_{1\mathbf{k}}+\mathcal{E}^2_{2\mathbf{k}}+\mathcal{E}^2_{3\mathbf{k}}}$ is the dispersion of the impurity subgap band, and

\begin{equation}
f_{\bk} = \sum_{m,n,s}\frac{\epsilon^{mns}}{2} \left[\mathcal{J}_{mx\bk}\mathcal{J}_{ny\bk}-\mathcal{J}_{my\bk}\mathcal{J}_{nx\bk}\right]\mathcal{E}_{s\bk} \,,
\end{equation}
where $\epsilon^{mns}$ denotes the Levi-Civita tensor with indices $m,n,s=1,2,3$. Obviously, $\sigma_\text{H}$ vanishes for any underlying even-parity chiral pairing, as their current operators contain only $\mathcal{J}_{3x(y)}$, even when further neighbor hoppings are included. In contrast, odd-parity pairings shall in general see a finite Hall conductance. This distinction applies to any superlattice configuration with no sublattice degree of freedom, including triangular lattices (see Table \ref{table1} and Ref.~\cite{supp}).

There are several features worth remarking. Firstly, the magnitude of $\sigma_\text{H}$ is determined by the above-defined hopping integrals which describe the hybridization between the impurity-bound states. Since these parameters overall grow exponentially with decreasing impurity spacing, one expects the conductivity to enhance exponentially with increasing impurity concentration. This contrast with the skew-scattering diagrammatic analysis which captures only the continuum state contributions~\cite{Goryo:08,Lutchyn:09}. Secondly, the cutoff frequency at which the imaginary part of $\sigma_\text{H}$ becomes nonzero (where a sharp peak appears) is set by the gap between the impurity bands. By contrast, continuum state contributions cuts off at $w=2\Delta_0$~\cite{Goryo:08,Lutchyn:09}. Finally, unlike the proposals which require particle-hole asymmetric normal state electron dispersion to obtain finite $\sigma_\text{H}$~\cite{Lutchyn:09,Taylor:12}, our low-energy theory has no such restriction.

{\it Honeycomb superlattice}-- The mixed particle-hole nature of the impurity subgap bands implies that there exists no fundamental symmetry constraints to prohibit the Hall effect in our effective theory. In other words, the vanishing of $\sigma_\text{H}$ in some of the models above must be accidental. Given that those models are characterized by single-sigma-matrix current operators, looking for systems that exhibit more structured current operators may be a promising route to obtain finite $\sigma_\text{H}$. One possibility is to introduce sublattice degrees of freedom. We verify this conjecture through a honeycomb lattice model [Fig.~\ref{fig1} (b)].

Consider up to nearest-neighbor terms, in the basis $\Psi_i=(c_{i,+},c^\prime_{i,+},c_{i,-},c^\prime_{i,-})^{\mathsf{T}}$ where $c$ and $c^\prime$ represent the two sublattices, the emergent tight-binding Hamiltonian has the form~\cite{supp},
\begin{equation}
\hat{H}^{\text{eff}}_{\bk}=
\begin{bmatrix}
    E_0 & \lambda_{\bk} & 0 & \eta_{\bk}\\
    \lambda_{\bk}^{*} & E_0 & (-1)^l\eta_{-\bk} & 0\\
    0 & (-1)^l\eta_{-\bk}^{*}& -E_0 &-\lambda_{\bk} \\
    \eta_{\bk}^{*} & 0 &-\lambda_{\bk}^{*} & -E_0
\end{bmatrix}
\end{equation}
where $\lambda_{\bk}=\sum_{\delta}e^{i\bk\cdot\bar\bR_{\delta}}\lambda$ and $\eta_{\bk}=\sum_{\delta}e^{i\bk\cdot\bar\bR_{\delta}}e^{il\theta_{\bar\bR_{\delta}}}\eta$, and $\bar\bR_{\delta}~(\delta=1,2,3)$ designate the three shortest vectors connecting sublattice $c$ to $c^\prime$. Interestingly, at $E_0=0$, the model resembles a low-energy theory proposed for the Moir\'e superlattice in twisted bilayer graphene~\cite{Yuan:18}.

As we have seen, in the case of even-parity pairing, the hopping between the `$+$' and `$-$' states on different sites does not generate particle current. However, the inter-sublattice hopping between the `$+$' (or `$-$') states introduces two off-diagonal components in the current operators. For example, in the present model,
\begin{equation}
\hat{J}^\text{eff}_{x\bk}=\mathcal{J}_{1x\bk}\varrho_1\otimes\sigma_3 +\mathcal{J}_{2x\bk}\varrho_2\otimes\sigma_3,
\end{equation}
in which $\varrho_i~(i=1,2,3)$ are the Pauli matrices operating in the sublattice manifold, and $\mathcal{J}_{1x\bk}=-3(\lambda^e-\lambda^h)\sin(\frac{3k_x}{2}) \cos(\frac{\sqrt{3}k_y}{2})$ and $\mathcal{J}_{2x\bk}=3(\lambda^e-\lambda^h)\cos(\frac{3k_x}{2}) \cos(\frac{\sqrt{3}k_y}{2})$. A lengthy calculation for $\sigma_\text{H}(w)$ presented in the Supplementary~\cite{supp} leads to an integral form involving
$[\mathcal{J}_{1x\bk}\mathcal{J}_{2y\bk}-\mathcal{J}_{2x\bk}\mathcal{J}_{1y\bk}] (|\eta_{\bk}|^2-|\eta_{-\bk}|^2)E_0$ in the numerator of the integrand. The integral is generically finite, in contrast to the square and triangular superlattice scenarios. For odd-parity pairings, an additional contribution to the current operators arises from the inter-sublattice hopping between the `$+$' and `$-$' states, and the Hall conductance is again finite.

\begin{table}[htbp]
    \centering
    \caption{Anomalous Hall effect in representative continuum chiral superconductors generated by impurity subgap bands emerging from various embedding impurity configurations. In comparison, the last row presents the prediction of the diagrammatic skew-scattering approach, which only considers the continuum state contribution~\cite{Goryo:08,Lutchyn:09}. }
    \begin{tabular}{c|c|c|c|c}
        \hline
         superlattice structure & p-wave & d-wave & f-wave& g-wave\\
         ~~ & ($l=1$) & ($l=2$) &($l=3$) & ($l=4$)\\
        \hline
        Square & \checkmark & $\times$ & \checkmark & $\times$ \\
        \hline
        Triangular & \checkmark & $\times$ & \checkmark & $\times$\\
        \hline
        Honeycomb & \checkmark & \checkmark & \checkmark & \checkmark \\
        \hline
      Random (continuum) & \checkmark & $\times$ & $\times$ & $\times$\\
        \hline
    \end{tabular}
    \label{table1}
\end{table}

{\it Concluding remarks}-- Table \ref{table1} summarizes our main results and makes a comparison to the conclusion obtained from the skew-scattering diagrammatic calculations~\cite{Goryo:08}. Since the latter approach had only accounted for contributions from the continuum states, our theory suggests that random impurities still have the potential to induce a finite Hall response in higher-order chiral superconducting states if the impurity-induced subgap states are considered. 

Current scanning tunnelling microscopy techniques have enabled atomically-controlled defect engineering~\cite{Khaj:19}, paving the way for studying the Hall effect in chiral superconductors with any desired impurity configuration. Our theory also suggests a viable means to probe charge-neutral chiral superfluids in liquid Helium-3~\cite{Anderson:61,Anderson:73} and in trapped cold Fermi gases~\cite{Liu:14,Buhler:14,Wu:16}. In these systems, ordered defects could be prepared using a periodically modulated holder potential or by shining patterned laser beams. In the case of quantum gases, for example, the resultant conductivity may be obtained by measuring the response of the system to a time-dependent trapping potential~\cite{Wu:15,Midtgaard:20}.

In summary, we have provided an alternative perspective on the impurity-induced anomalous Hall response in chiral superconductors. Previous studies of this kind had relied exclusively on diagrammatic approach involving extensive vertex correction analyses, and they had only accounted for the contribution from the quasiparticle continuum. By laying our eyes on the effect of impurity induced bound states, we constructed an emergent low energy theory for when the impurities are deposited in superlattice patterns. Owing to the mixed particle-hole character of the bound states, the resultant theory generates finite Hall conductivity at the one-loop level of the approximation - without the need for vertex corrections. 

{\it Acknowledgements:} We acknowledge fruitful discussions with Fu-Chun Zhang. This work is supported by NSFC under grant No.~11904155 (WH), the Guangdong Provincial Key Laboratory under Grant No.~2019B121203002 (WH), and the China Postdoctoral Science Foundation under Grant No. 2020M670422 (YL).

\newpage

\setcounter{figure}{0}
\setcounter{equation}{0}
\renewcommand {\theequation} {S\arabic{equation}}
\renewcommand {\thefigure} {S\arabic{figure}}
\widetext
\begin{center}
\textbf{\large Supplemental Material for ``Anomalous Hall effect in single-band chiral superconductors from impurity superlattices"}\\
\vspace{4mm}
\author{Yu Li}
\address{Shenzhen Institute for Quantum Science and Engineering \& Guangdong Provincial Key Laboratory of Quantum Science and Engineering, Southern University of Science and Technology, Shenzhen 518055, Guangdong, China}
\affiliation{Kavli Institute for Theoretical Sciences, University of Chinese Academy of Sciences, Beijing 100049, China}
\author{Zhiqiang Wang}
\affiliation{Department of Physics and Astronomy, McMaster University, Hamilton, Ontario L8S 4M1, Canada}
\affiliation{James Franck Institute, University of Chicago, Chicago, Illinois 60637, USA}
\author{Wen Huang}
\email{huangw3@sustech.edu.cn}
\address{Shenzhen Institute for Quantum Science and Engineering \& Guangdong Provincial Key Laboratory of Quantum Science and Engineering, Southern University of Science and Technology, Shenzhen 518055, Guangdong, China}
\date{\today}
\end{center}

\section{I. Impurity-induced bound states in chiral superconductors}

The BdG Hamiltonian for a two-dimensional chiral
superconductor with impurities can be written as%
\begin{equation}
H_{\text{BdG}}=H_{\text{BdG}}^{\left(  \text{bulk}\right)  }+H^{\left(
\text{imp}\right)  },
\end{equation}
which is expressed in the Nambu space spanned by the spinor $\phi_{\mathbf{k}}=(
c_{\mathbf{k}},c_{-\mathbf{k}}^{\dagger})^{\intercal}$, where $c_{\mathbf{k}}$
$(  c_{\mathbf{k}}^{\dagger})$ is the electron annihilation
(creation) operator. In the continuum limit, the bulk
Hamiltonian in the momentum space, $H^{\left(  \text{bulk}\right)  }_{\mathbf{k}}$, can be
expanded in terms of the Pauli matrices $\tau_{i}$ ($i=1,2,3$) as%
\begin{equation}
H_{\text{BdG}}^{\left(  \text{bulk}\right)  }=\int\frac{d\mathbf{k}}{\left(
2\pi\right)  ^{2}}H_{\mathbf{k}}^{\left(  \text{bulk}\right)  }=\int%
\frac{d\mathbf{k}}{\left(  2\pi\right)  ^{2}}\left[  \epsilon_{\mathbf{k}}%
\tau_{3}+\operatorname{Re}\left(  \Delta_{\mathbf{k}}\right)  \tau
_{1}-\operatorname{Im}\left(  \Delta_{\mathbf{k}}\right)  \tau_{2}\right]  .
\end{equation}
$\epsilon_{\mathbf{k}}=\mathbf{k}^{2}/2m-\varepsilon
_{\text{F}}$ is the dispersion of electrons relative to the Fermi energy $\varepsilon_{\text{F}}$. $\Delta_{\mathbf{k}}=\Delta e^{il\theta_{\mathbf{k}}}%
$ is the gap function of the chiral pairing, where $\theta_{\mathbf{k}}$ is the azimuthal angle of
$\mathbf{k}$, and $l$ and $\Delta$ represent, respectively, the Cooper pair angular quantum
number and the $k$ independent gap magnitude.

We first solve a single-impurity problem, with a delta-function potential with strength $U$ located at the origin,
\begin{equation}
H^{\left(  \text{imp}\right)  }\left(  \mathbf{r}\right)  =U\tau_{3}\delta\left(  \mathbf{r}\right).
\end{equation}
The equation to be solved is
$H_{\text{BdG}}\psi\left(  \mathbf{r}\right)  =E\psi\left(  \mathbf{r}\right) $, where $E$ is the eigenvalue. Performing a Fourier transformation
$\psi\left(  \mathbf{r}\right)  =\int\frac{d\mathbf{k}}{\left(  2\pi\right)
^{2}}e^{i\mathbf{k\cdot r}}\psi_{\mathbf{k}}$ , one
obtains%
\begin{equation}
\left[  E-H_{\mathbf{k}}^{\left(  \text{bulk}\right)  }\right]
\psi_{\mathbf{k}}=U\tau_{3}\psi\left( 0\right)  .
\end{equation}
Transformed back into the real space, the wavefunction becomes%
\begin{equation}
\psi\left(  \mathbf{r}\right)  =UG\left(  E,\mathbf{r}\right)
\tau_{3}\psi\left(  0\right)  ,
\end{equation}
where $G\left(  E,\mathbf{r}\right)  $ is the bulk Green's function, %
\begin{align} \label{eq:Green}
G\left(  E,\mathbf{r}\right)   &  =\int\frac{d\mathbf{k}}{\left(
2\pi\right)  ^{2}}e^{i\mathbf{k\cdot r}}\left[  E-H_{\mathbf{k}}^{\left(
\text{bulk}\right)  }\right]  ^{-1}=\int\frac{d\mathbf{k}}{\left(
2\pi\right)  ^{2}}e^{i\mathbf{k\cdot r}}\frac{E\tau_{0} + \epsilon
_{\mathbf{k}}\tau_{3} + \operatorname{Re}\left(  \Delta_{\mathbf{k}}\right)
\tau_{1} - \operatorname{Im}\left(  \Delta_{\mathbf{k}}\right)  \tau_{2}}%
{E^{2}-\epsilon_{\mathbf{k}}^{2}-\left\vert \Delta_{\mathbf{k}}\right\vert
^{2}}\nonumber\\
&  =X_{0}\tau_{0}+X_{1}\tau_{3}+iX_{2}^{+}\tau_{+}+iX_{2}^{-}\tau_{-},
\end{align}
in which $\tau_{\pm}=\left(  \tau_{1}\pm i\tau_{2}\right)  /2$, and $X_{0}$,
$X_{1}$, $X_{2}^{\pm}$ are given by~\cite{Kaladzhyan16a,Kaladzhyan16b},
\begin{equation}
X_{0}\left(  E,\mathbf{r}\right)  = - \int\frac{d\mathbf{k}}{\left(
2\pi\right)  ^{2}}\frac{Ee^{i\mathbf{k\cdot r}}}{\epsilon_{\mathbf{k}}%
^{2}+\left\vert \Delta_{\mathbf{k}}\right\vert ^{2}-E^{2}}
\approx
-\frac{2N_{\text{F}}E}{\sqrt{\Delta^2-E^{2}}%
}\operatorname{Im}\left\{  K_{0}\left[  \left(  \kappa-i\right)
k_{\text{F}}r\right]  \right\}  ,
\label{eq:X0}
\end{equation}%
\begin{equation}
X_{1}\left(  E,\mathbf{r}\right)  =-\int\frac{d\mathbf{k}}{\left(
2\pi\right)  ^{2}}\frac{\epsilon_{\mathbf{k}}e^{i\mathbf{k\cdot r}}}%
{\epsilon_{\mathbf{k}}^{2}+\left\vert \Delta_{\mathbf{k}}\right\vert
^{2}-E^{2}}
\approx
-2N_{\text{F}}\operatorname{Re}\left\{ K_{0}\left[  \left( \kappa-i\right)  k_{\text{F}}r\right]  \right\}
,
\label{eq:X1}
\end{equation}%
\begin{equation}
X_{2}^{\pm}\left(  E,\mathbf{r}\right)  =\pm\int\frac{d\mathbf{k}}{\left(
2\pi\right)  ^{2}}\frac{i \Delta e^{\pm il\theta_{\mathbf{k}}}%
e^{i\mathbf{k\cdot r}}}{\epsilon_{\mathbf{k}}^{2}+\left\vert \Delta
_{\mathbf{k}}\right\vert ^{2}-E^{2}}
\approx
\pm e^{\pm i (l+1)\pi/2}
\frac{2N_{\text{F}} \Delta}{\sqrt{\Delta^2 -E^2}}
e^{\pm il\theta_{\mathbf{r}}}
\operatorname{Im}%
\left\{
K_{l}\left[  \left( \kappa-i\right)  k_{\text{F}}r\right]
\right\} \,.
\label{eq:X2}
\end{equation}
In these expressions, $N_{\text{F}}$ is the density of states at $\epsilon_{\text{F}}$,
$\kappa \equiv \sqrt{\Delta^2-E^2}/(k_{\text{F}} \upsilon_{\text{F}})$, and the function $K_{n}\left(
x\right)  $ represents the modified Bessel functions of
the second kind of order $n$.
Far from the impurity, $X_{0}$, $X_{1}$, $X_{2}^{\pm}$ all decay as $e^{-r/\xi
}/\sqrt{k_{\text{F}}r}$, in which $\xi=\upsilon_{\text{F}}/\sqrt{\Delta^2-E^2}$ is the effective coherence length.
The above equations are valid for $\left\vert E\right\vert <\Delta$ and $\kappa k_{\text{F}}r=r/\xi \gtrsim 1$.
Note that the Bessel functions involved diverge at $r=0$. Right at $r=0$, the $\mathbf{k}$ integral can be performed without resorting to
Bessel functions, leading to
\begin{equation}
X_0(E,0)= -\frac{\pi N_F E}{\sqrt{\Delta^2-E^2}} ,  X_1(E,0)=0 ,  X_2^{\pm}(E,0)=0.
\label{eq:Xat0}
\end{equation}
 An ultraviolet energy cut off is needed to regulate the divergence in the $\mathbf{k}$ integrals,
in order to obtain the correct behavior of the Green's function at $0<r/\xi \lesssim 1$. However, we will ignore this short-distance behavior since it is not important for our
following discussions.


At $\mathbf{r =0}$, i.e., right at the impurity site, the eigenvalue equation becomes
\begin{equation}
\left[  1-UG\left(  E,0\right)  \tau_{3}\right]  \psi\left( 0\right)  =0.
\end{equation}
Using Eqs.~\eqref{eq:Xat0} and \eqref{eq:Green} we obtain the impurity induced subgap state energies as $E=\pm E_0$
with $E_0=\Delta/\sqrt{1+\beta^2}$, where $\beta=\pi N_{\text{F}}U$. The two energies are symmetric with respect to $E=0$, which is not the case in general
if the particle-hole asymmetry of the normal state energy dispersion is introduced; also, the expression of $E_0$ is independent
of the sign of $U$, which needs to be modified if the $k$-dependence of the gap function is included. However,  considering more general cases does not alter the conclusions obtained in the main text. We denote the two eigenvectors corresponding to $E=\pm E_0$ as
$\psi_{+}\left(  \mathbf{r}\right)  $ and $\psi_{-}\left(  \mathbf{r}\right)$, respectively,
and consider the $U>0$ (repulsive)and $U<0$ (attractive) cases separately in the following. 
\begin{enumerate}[(1)]
\item For  $U>0$ the two eigenvectors at $\mathbf{r =0}$ are $\psi_{+}(0)
=(1,0)  ^{\intercal}$ (particle-like) and $\psi_{-}(0)  =\left(  0,1\right)  ^{\intercal}$ (hole-like).
At $\mathbf{r \ne 0}$
\begin{align}
\psi_{+}\left(  \mathbf{r}\right)  
& =
 \frac{1}{\mathcal{N}} UG\left(  +E_0,\mathbf{r}\right)
\tau_{3}\psi_+ \left( 0\right) 
\nonumber \\
& =
\frac{1}{\sqrt{
\left[
X_0\left( E_0,\mathbf{r}\right)+X_1\left(E_0,\mathbf{r}\right)
\right]^2
+\vert X_2^{+}\left( E_0,\mathbf{r}\right)\vert^2}}\left(
\begin{array}
[c]{c}%
X_{0}\left( E_0,\mathbf{r}\right)+X_{1}\left(E_0,\mathbf{r}\right)\\
iX_{2}^{-}\left(E_0,\mathbf{r}\right)%
\end{array}
\right)  \equiv\left(
\begin{array}
[c]{c}%
u\left(  \mathbf{r}\right) \\
\upsilon\left(  \mathbf{r}\right)
\end{array}
\right)  ,
\end{align}%
where $\mathcal{N}$ is a normalization coefficient and, similarly,
\begin{equation}
\psi_{-}\left(  \mathbf{r}\right)  
=
\frac{1}{\sqrt{
\left[
-X_0\left( -E_0,\mathbf{r}\right)+X_1\left(-E_0,\mathbf{r}\right)
\right]^2
+\vert X_2^{+}\left( -E_0,\mathbf{r}\right)\vert^2}}
\left(
\begin{array}
[c]{c}%
-iX_{2}^{+}\left(  -E_0,\mathbf{r}\right)\\
-X_{0}\left(  -E_0,\mathbf{r}\right)+X_{1}\left(  -E_0,\mathbf{r}\right)%
\end{array}
\right)  =\left(
\begin{array}
[c]{c}%
-\upsilon^{\ast}\left(  \mathbf{r}\right) \\
u^{\ast}\left(  \mathbf{r}\right)
\end{array}
\right)  .
\end{equation}
Note that $X_{0}\left( E,\mathbf{r}\right)$ is odd in $E$, while $X_{1}\left( E,\mathbf{r}\right)$ and $X_{2}^{\pm}\left( E,\mathbf{r}\right)$ are both even in $E$. 
From Eqns \ref{eq:X0}-\ref{eq:X2} we see that $u\left(  \mathbf{r}\right)$ is real for the given $\psi_+(0)$ and $\psi_-(0)$, and we can write $\psi_{+}\left(
\mathbf{r}\right)  =\left(  u\left(  \mathbf{r}\right)  ,\upsilon\left(
\mathbf{r}\right)  \right)  ^{\intercal}=$ $\left(  u_{r},e^{-il\theta
_{\mathbf{r}}+\alpha}\upsilon_{r}\right)  ^{\intercal}$, where $u_{r}$ and
$\upsilon_{r}$ are two real functions of $r$ only, 
and $\alpha$ is an $\mathbf{r}$-independent phase. For notational simplicity, we will set $\alpha=0$, which will not qualitatively affect our conclusions. 

\item The eigenvectors for $U<0$ can be obtained similarly. At $\mathbf{r=0}$, $\psi_{+}(0)
=(0,1)^{\intercal}$ (hole-like) and $\psi_{-}(0)  =\left( 1,0\right)  ^{\intercal}$ (particle-like). At $\mathbf{r \neq 0}$,
\begin{equation}
\psi_{+}\left(  \mathbf{r}\right)  
=\frac{1}{\sqrt{[-X_0\left( E_0,\mathbf{r}\right)+X_1\left( E_0,\mathbf{r}\right)]^2+|X_2^{+}\left( E_0,\mathbf{r}\right)|^2}}\left(
\begin{array}
[c]{c}%
-iX_{2}^{+}\left(E_0,\mathbf{r}\right)\\
-X_{0}\left(E_0,\mathbf{r}\right)+X_{1}\left(E_0,\mathbf{r}\right)%
\end{array}
\right)  \equiv\left(
\begin{array}
[c]{c}%
-\upsilon^{\prime\ast}\left(  \mathbf{r}\right) \\
u^{\prime\ast}\left(  \mathbf{r}\right)
\end{array}
\right)  ,
\end{equation}%
\begin{equation}
\psi_{-}\left(  \mathbf{r}\right)  
=\frac{1}{\sqrt{[X_0\left( -E_0,\mathbf{r}\right)+X_1\left( -E_0,\mathbf{r}\right)]^2+|X_2^{+}\left( -E_0,\mathbf{r}\right)|^2}}\left(
\begin{array}
[c]{c}%
X_{0}\left(  -E_0,\mathbf{r}\right)+X_{1}\left(  -E_0,\mathbf{r}\right)\\
iX_{2}^{-}\left(  -E_0,\mathbf{r}\right)%
\end{array}
\right)  =\left(
\begin{array}
[c]{c}%
u^{\prime}\left(  \mathbf{r}\right) \\
\upsilon^{\prime}\left(  \mathbf{r}\right)
\end{array}
\right)  .
\end{equation}
Again, $u^{\prime}(\br)$ is real, and we can write $\psi_{-}\left(
\mathbf{r}\right)  =\left(  u^{\prime}\left(  \mathbf{r}\right)  ,\upsilon^{\prime}\left(\mathbf{r}\right)  \right)  ^{\intercal}=$ $\left(  u^{\prime}_{r},e^{-il\theta
_{\mathbf{r}}+ i \alpha}\upsilon^{\prime}_{r}\right)  ^{\intercal}$, where $u^{\prime}_{r}$ and
$\upsilon^{\prime}_{r}$ are real functions of $r$, and $\alpha$ is again a constant phase we will set to be zero without altering our conclusions.
\end{enumerate}
In the main text and in the following discussions, we only consider the case with repulsive $U$. The attractive-$U$ scenario produces similar physics.

\section{II. Low-energy effective model of the impurity superlattice}
In an impurity lattice, the bound states from different impurity sites hybridize through the kinetic hopping and Cooper pairing in the original microscopic BdG Hamiltonian, forming subgap bands. Treating the `$+$' and `$-$' bound states on each impurity site as two independent orbitals, we now construct an effective tight-binding Hamiltonian for the subgap states on an impurity lattice. In the second-quantization formulation, the creation (annihilation) of the orbitals are denoted by the operators $c_{\pm}^{\dagger}$ ($c_{\pm}$). We first consider a two-impurity system with impurities located at $\mathbf{R}_{i}$ and $\mathbf{R}_{j}$. In the basis $\hat{\Psi}_{i}=\left(  c_{i,+}%
,c_{i,-}\right)  ^{\intercal}$ where $i$ is the site index, the emergent effective Hamiltonian reads $H=\sum_{i,j}\hat{\Psi}_{i}^{\dagger}\left[  E_{0}\delta
_{ij}\sigma_{3}+\hat{h}_{ij}\left(  1-\delta_{ij}\right)  \right]  \hat{\Psi
}_{j}+$H.c., in which
\begin{equation}
h_{ij}=\left(
\begin{array}
[c]{cc}%
t_{ij}^{++} & t_{ij}^{+-}\\
t_{ij}^{-+} & t_{ij}^{--}%
\end{array}
\right)  ,
\end{equation}
where%
\begin{equation}
t_{ij}^{\mu\nu}=\int d\mathbf{r}d\mathbf{r}^{\prime}\psi_{\mu}^{\dagger
}\left(  \mathbf{r-R}_{i}\right)  H_{\text{BdG}}^{\left(  \text{bulk}\right)
}\left(  \mathbf{r,r}^{\prime}\right)  \psi_{\nu}\left(  \mathbf{r}^{\prime
}\mathbf{-R}_{j}\right)  ,
\end{equation}
and $\mu,\nu=+,-$. Explicitly,%
\begin{align}
t_{ij}^{++}  =& \int d\mathbf{r}d\mathbf{r}^{\prime}\left\{  u_{\left\vert
\mathbf{r-R}_{i}\right\vert }\left[  \delta_{\mathbf{r,r}^{\prime}}\left(
-\frac{\nabla_{\mathbf{r}^{\prime}}^{2}}{2m_{e}}-\mu\right)  \right]
u_{\left\vert \mathbf{r}^{\prime}\mathbf{-R}_{j}\right\vert }+u_{\left\vert
\mathbf{r-R}_{i}\right\vert }\Delta\left(  \mathbf{r-r}^{\prime}\right)
e^{-il\theta_{\mathbf{r}^{\prime}\mathbf{-R}_{j}}}\upsilon_{\left\vert
\mathbf{r}^{\prime}\mathbf{-R}_{j}\right\vert }\right. \nonumber\\
&  \left.  +e^{il\theta_{\mathbf{r-R}_{i}}}\upsilon_{\left\vert \mathbf{r-R}%
_{i}\right\vert }\Delta^{\ast}\left(  \mathbf{r-r}^{\prime}\right)
u_{\left\vert \mathbf{r}^{\prime}\mathbf{-R}_{j}\right\vert }+e^{il\theta
_{\mathbf{r-R}_{i}}}\upsilon_{\left\vert \mathbf{r-R}_{i}\right\vert }\left[
\delta_{\mathbf{r}^{\prime}\mathbf{,r}}\left(  \frac{\nabla_{\mathbf{r}%
^{\prime}}^{2}}{2m_{e}}+\mu\right)  \right]  e^{-il\theta_{\mathbf{r}^{\prime
}\mathbf{-R}_{j}}}\upsilon_{\left\vert \mathbf{r}^{\prime}\mathbf{-R}%
_{j}\right\vert }\right\}  ,\\
t_{ij}^{+-}  = & \int d\mathbf{r}d\mathbf{r}^{\prime}\left\{  -u_{\left\vert
\mathbf{r-R}_{i}\right\vert }\left[  \delta_{\mathbf{r,r}^{\prime}}\left(
-\frac{\nabla_{\mathbf{r}^{\prime}}^{2}}{2m_{e}}-\mu\right)  \right]
e^{il\theta_{\mathbf{r}^{\prime}\mathbf{-R}_{j}}}\upsilon_{\left\vert
\mathbf{r}^{\prime}\mathbf{-R}_{j}\right\vert }+u_{\left\vert \mathbf{r-R}%
_{i}\right\vert }\Delta\left(  \mathbf{r-r}^{\prime}\right)  u_{\left\vert
\mathbf{r}^{\prime}\mathbf{-R}_{j}\right\vert }\right. \nonumber\\
&\left.  -e^{il\theta_{\mathbf{r-R}_{i}}}\upsilon_{\left\vert \mathbf{r-R}%
_{i}\right\vert }\Delta^{\ast}\left(  \mathbf{r-r}^{\prime}\right)
e^{il\theta_{\mathbf{r}^{\prime}\mathbf{-R}_{j}}}\upsilon_{\left\vert
\mathbf{r}^{\prime}\mathbf{-R}_{j}\right\vert }+e^{il\theta_{\mathbf{r-R}_{i}%
}}\upsilon_{\left\vert \mathbf{r-R}_{i}\right\vert }\left[  \delta
_{\mathbf{r}^{\prime}\mathbf{,r}}\left(  \frac{\nabla_{\mathbf{r}^{\prime}%
}^{2}}{2m_{e}}+\mu\right)  \right]  u_{\left\vert \mathbf{r}^{\prime
}\mathbf{-R}_{j}\right\vert }\right\}  ,
\end{align}
\begin{align}
t_{ij}^{-+}  =& \int d\mathbf{r}d\mathbf{r}^{\prime}\left\{  -e^{-il\theta
_{\mathbf{r-R}_{i}}}\upsilon_{\left\vert \mathbf{r-R}_{i}\right\vert }\left[
\delta_{\mathbf{r,r}^{\prime}}\left(  -\frac{\nabla_{\mathbf{r}^{\prime}}^{2}%
}{2m_{e}}-\mu\right)  \right]  u_{\left\vert \mathbf{r}^{\prime}%
\mathbf{-R}_{j}\right\vert }-e^{-il\theta_{\mathbf{r-R}_{i}}}\upsilon
_{\left\vert \mathbf{r-R}_{i}\right\vert }\Delta\left(  \mathbf{r-r}^{\prime
}\right)  e^{-il\theta_{\mathbf{r}^{\prime}\mathbf{-R}_{j}}}\upsilon
_{\left\vert \mathbf{r}^{\prime}\mathbf{-R}_{j}\right\vert }\right.
\nonumber\\
&  \left.  +u_{\left\vert \mathbf{r-R}_{i}\right\vert }\Delta^{\ast}\left(
\mathbf{r-r}^{\prime}\right)  u_{\left\vert \mathbf{r}^{\prime}\mathbf{-R}%
_{j}\right\vert }+u_{\left\vert \mathbf{r-R}_{i}\right\vert }\left[
\delta_{\mathbf{r}^{\prime}\mathbf{,r}}\left(  \frac{\nabla_{\mathbf{r}%
^{\prime}}^{2}}{2m_{e}}+\mu\right)  \right]  e^{-il\theta_{\mathbf{r}^{\prime
}\mathbf{-R}_{j}}}\upsilon_{\left\vert \mathbf{r}^{\prime}\mathbf{-R}%
_{j}\right\vert }\right\}  ,\\
t_{ij}^{--} = &\int d\mathbf{r}d\mathbf{r}^{\prime}\left\{  e^{-il\theta
_{\mathbf{r-R}_{i}}}\upsilon_{\left\vert \mathbf{r-R}_{i}\right\vert }\left[
\delta_{\mathbf{r,r}^{\prime}}\left(  -\frac{\nabla_{\mathbf{r}^{\prime}}^{2}%
}{2m_{e}}-\mu\right)  \right]  e^{il\theta_{\mathbf{r}^{\prime}\mathbf{-R}%
_{j}}}\upsilon_{\left\vert \mathbf{r}^{\prime}\mathbf{-R}_{j}\right\vert
}-e^{-il\theta_{\mathbf{r-R}_{i}}}\upsilon_{\left\vert \mathbf{r-R}%
_{i}\right\vert }\Delta\left(  \mathbf{r-r}^{\prime}\right)  u_{\left\vert
\mathbf{r}^{\prime}\mathbf{-R}_{j}\right\vert }\right. \nonumber\\
&  \left.  -u_{\left\vert \mathbf{r-R}_{i}\right\vert }\Delta^{\ast}\left(
\mathbf{r-r}^{\prime}\right)  e^{il\theta_{\mathbf{r}^{\prime}\mathbf{-R}_{j}%
}}\upsilon_{\left\vert \mathbf{r}^{\prime}\mathbf{-R}_{j}\right\vert
}+u_{\left\vert \mathbf{r-R}_{i}\right\vert }\left[  \delta_{\mathbf{r}%
^{\prime}\mathbf{,r}}\left(  \frac{\nabla_{\mathbf{r}^{\prime}}^{2}}{2m_{e}%
}+\mu\right)  \right]  u_{\left\vert \mathbf{r}^{\prime}\mathbf{-R}%
_{j}\right\vert }\right\}  .
\end{align}
From these expressions, one can easily obtain the relations, $t_{ij}%
^{++}=-\left(  t_{ij}^{--}\right)  ^{\ast}\equiv\lambda_{ij}$, $t_{ij}%
^{+-}=\left(  t_{ij}^{-+}\right)  ^{\dagger}\equiv\eta_{ij}$. The hybridization has three distinct origins: electron-electron hopping, hole-hole
hopping and Cooper pairing. Hence we decompose the hopping terms as $\lambda_{ij}=\lambda_{ij}^{e}+\lambda_{ij}^{h}+\lambda_{ij}^{\Delta}$, and $\eta_{ij}=\eta_{ij}^{e}+\eta_{ij}%
^{h}+\eta_{ij}^{\Delta}$, the details of which we provide below.

\subsection{Symmetry aspects of the hybridization matrix elements}

By changing the variables, one can easily find that $\lambda_{ij}$ and $\eta_{ij}$ depend on the relative position of $\mathbf{R}_{i}$ and $\mathbf{R}_{j}$, i.e.,
$\lambda_{ij}\equiv \lambda\left(  \mathbf{R}_{j}-\mathbf{R}_{i}\right)  $,
$\eta_{ij}\equiv \eta\left(  \mathbf{R}_{j}-\mathbf{R}_{i}\right)  $. Define $  \mathbf{R}_{\delta}=\mathbf{R}_{j}-\mathbf{R}_{i}$, the expressions for $\lambda\left(  \mathbf{R}_{\delta}\right)$ and $\eta\left(  \mathbf{R}_{\delta}\right)$ can be reduced as

\begin{align}
\lambda\left(  \mathbf{R}_{\delta}\right)   =& \lambda^{e}\left(
\mathbf{R}_{\delta}\right)  +\lambda^{h}\left(  \mathbf{R}_{\delta}\right)
+\lambda^{\Delta}\left(  \mathbf{R}_{\delta}\right) \nonumber\\
=&\int d\mathbf{r}d\mathbf{r}^{\prime}\left\{  u_{r}\left[  \delta
_{\mathbf{r,r}^{\prime}-\mathbf{R}_{\delta}}\left(  -\frac{\nabla
_{\mathbf{r}^{\prime}}^{2}}{2m_{e}}-\mu\right)  \right]  u_{r^{\prime}%
}+e^{il\theta_{\mathbf{r}}}\upsilon_{r}\left[  \delta_{\mathbf{r}^{\prime
}-\mathbf{R}_{\delta}\mathbf{,r}}\left(  \frac{\nabla_{\mathbf{r}^{\prime}%
}^{2}}{2m_{e}}+\mu\right)  \right]  e^{-il\theta_{\mathbf{r}^{\prime}}}\upsilon
_{r^{\prime}}\right. \nonumber\\
&  \left.  +2\operatorname{Re}\left[  \Delta\left(  \mathbf{r-r}^{\prime
}-\mathbf{R}_{\delta}\right)  e^{-il\theta_{\mathbf{r}^{\prime}}}\right]
u_{r}\upsilon_{r^{\prime}}\right\}  ,\\
\nonumber\\
\eta\left(  \mathbf{R}_{\delta}\right)  =  &  \eta^{e}\left(  \mathbf{R}%
_{\delta}\right)  +\eta^{h}\left(  \mathbf{R}_{\delta}\right)  +\eta^{\Delta
}\left(  \mathbf{R}_{\delta}\right) \nonumber\\
=  &  \int d\mathbf{r}d\mathbf{r}^{\prime}\left\{  \left(  1+e^{il\pi}\right)
\int d\mathbf{r}d\mathbf{r}^{\prime}u_{r}\delta_{\mathbf{r,r}^{\prime
}-\mathbf{R}_{\delta}}\left(  \frac{\nabla_{\mathbf{r}^{\prime}}^{2}}{2m_{e}%
}+\mu\right)  e^{il\theta_{\mathbf{r}^{\prime}}}\upsilon_{r^{\prime}}%
+\Delta\left(  \mathbf{r-r}^{\prime}-\mathbf{R}_{\delta}\right)
u_{r}u_{r^{\prime}}\right. \nonumber\\
&  \left.  -e^{il\left(  \theta_{\mathbf{r}}+\theta_{\mathbf{r}^{\prime}%
}\right)  }\Delta^{\ast}\left(  \mathbf{r-r}^{\prime}-\mathbf{R}_{\delta
}\right)  \upsilon_{r}\upsilon_{r^{\prime}}\right\}  .
\end{align}

To inspect the dependence of $\lambda\left(  \mathbf{R}_{\delta}\right)  $ and
$\eta\left(  \mathbf{R}_{\delta}\right)  $ on the orientation of $ \mathbf{R}_{\delta}$, let us perform a rotation
($\hat{R}$) of arbitrary angle $\phi$. Then,%
\begin{align}
\lambda\left(  \hat{R}_{\phi}\mathbf{R}_{\delta}\right)   =&\int
d\mathbf{r}d\mathbf{r}^{\prime}\left\{  u_{r}\left[  \delta_{\mathbf{r,r}%
^{\prime}-\hat{R}_{\phi}\mathbf{R}_{\delta}}\left(  -\frac{\nabla
_{\mathbf{r}^{\prime}}^{2}}{2m_{e}}-\mu\right)  \right]  u_{r^{\prime}%
}+e^{il\theta_{\mathbf{r}}}\upsilon_{r}\left[  \delta_{\mathbf{r}^{\prime
}-\hat{R}_{\phi}\mathbf{R}_{\delta}\mathbf{,r}}\left(  \frac{\nabla
_{\mathbf{r}^{\prime}}^{2}}{2m_{e}}+\mu\right)  \right]  e^{-il\theta
_{r^{\prime}}}\upsilon_{r^{\prime}}\right. \nonumber\\
&  \left.  +2\operatorname{Re}\left[  \Delta\left(  \mathbf{r-r}^{\prime}%
-\hat{R}_{\phi}\mathbf{R}_{\delta}\right)  e^{-il\theta_{\mathbf{r}^{\prime}}%
}\right]  u_{r}\upsilon_{r^{\prime}}\right\} \nonumber\\
=&\int d\mathbf{r}d\mathbf{r}^{\prime}\left\{  u_{r}\left[  \delta_{\hat
{R}_{\phi}\mathbf{r,}\hat{R}_{\phi}\left(  \mathbf{r}^{\prime}-\mathbf{R}%
_{\delta}\right)  }\left(  -\frac{\nabla_{\hat{R}_{\phi}\mathbf{r}^{\prime}%
}^{2}}{2m_{e}}-\mu\right)  \right]  u_{r^{\prime}}+e^{il\theta_{\hat{R}_{\phi
}\mathbf{r}}}\upsilon_{r}\left[  \delta_{\hat{R}_{\phi}\left(  \mathbf{r}%
^{\prime}-\mathbf{R}_{\delta}\right)  \mathbf{,}\hat{R}_{\phi}\mathbf{r}%
}\left(  \frac{\nabla_{\hat{R}_{\phi}\mathbf{r}^{\prime}}^{2}}{2m_{e}}%
+\mu\right)  \right]  \right. \nonumber\\
&  \left.  \times e^{-il\theta_{\hat{R}_{\phi}r^{\prime}}}\upsilon_{r^{\prime
}}+2\operatorname{Re}\left[  \Delta\left(  \hat{R}_{\phi}\left(
\mathbf{r-r}^{\prime}-\mathbf{R}_{\delta}\right)  \right)  e^{-il\theta
_{\hat{R}_{\phi}\mathbf{r}^{\prime}}}\right]  u_{r}\upsilon_{r^{\prime}%
}\right\} \nonumber\\
=&\int d\mathbf{r}d\mathbf{r}^{\prime}\left\{  u_{r}\left[  \delta
_{\mathbf{r,}\left(  \mathbf{r}^{\prime}-\mathbf{R}_{\delta}\right)  }\left(
-\frac{\nabla_{\hat{R}_{\phi}\mathbf{r}^{\prime}}^{2}}{2m_{e}}-\mu\right)
\right]  u_{r^{\prime}}+e^{il\phi}e^{il\theta_{\mathbf{r}}}\upsilon_{r}\left[
\delta_{\hat{R}_{\phi}\left(  \mathbf{r}^{\prime}-\mathbf{R}_{\delta}\right)
\mathbf{,}\hat{R}_{\phi}\mathbf{r}}\left(  \frac{\nabla_{\hat{R}_{\phi
}\mathbf{r}^{\prime}}^{2}}{2m_{e}}+\mu\right)  \right]  \right. \nonumber\\
&  \left.  \times e^{-il\phi}e^{-il\theta_{r^{\prime}}}\upsilon_{r^{\prime}%
}+2\operatorname{Re}\left[  e^{il\phi}\Delta\left(  \mathbf{r-r}^{\prime
}-\mathbf{R}_{\delta}\right)  e^{-il\phi}e^{-il\theta_{\mathbf{r}^{\prime}}}\right]  u_{r}\upsilon_{r^{\prime}}\right\} \nonumber\\
=& \lambda\left(  \mathbf{R}_{\delta}\right)  ,
\end{align}%
which is independent of the orientation of $\mathbf{R}_{\delta}$, i.e., $\lambda\left(  \mathbf{R}_{\delta}\right)=\lambda\left( | \mathbf{R}_{\delta}|\right)$. And
\begin{align}
\eta\left(  \hat{R}_{\phi}\mathbf{R}_{\delta}\right)   =&\int d\mathbf{r}%
d\mathbf{r}^{\prime}\left\{  \Delta\left(  \mathbf{r-r}^{\prime}+\hat{R}%
_{\phi}\mathbf{R}_{\delta}\right)  u_{r}u_{r^{\prime}}-e^{il\left(
\theta_{\mathbf{r}}+\theta_{\mathbf{r}^{\prime}}\right)  }\Delta^{\ast}\left(
\mathbf{r-r}^{\prime}+\hat{R}_{\phi}\mathbf{R}_{\delta}\right)  \upsilon
_{r}\upsilon_{r^{\prime}}\right. \nonumber\\
&  \left.  +\left(  1+e^{il\pi}\right)  \int d\mathbf{r}d\mathbf{r}^{\prime
}u_{r}\left[  \delta_{\mathbf{r,r}^{\prime}+\hat{R}_{\phi}\mathbf{R}_{\delta}%
}\left(  \frac{\nabla_{\mathbf{r}^{\prime}}^{2}}{2m_{e}}+\mu\right)  \right]
e^{il\theta_{\mathbf{r}^{\prime}}}\upsilon_{r^{\prime}}\right\} \nonumber\\
=&\int d\mathbf{r}d\mathbf{r}^{\prime}\left\{  \Delta\left(  \hat{R}_{\phi
}\left(  \mathbf{r-r}^{\prime}+\mathbf{R}_{\delta}\right)  \right)
u_{r}u_{r^{\prime}}-e^{il\left(  \theta_{\hat{R}_{\phi}\mathbf{r}}%
+\theta_{\hat{R}_{\phi}\mathbf{r}^{\prime}}\right)  }\Delta^{\ast}\left(
\hat{R}_{\phi}\left(  \mathbf{r-r}^{\prime}+\mathbf{R}_{\delta}\right)
\right)  \upsilon_{r}\upsilon_{r^{\prime}}\right. \nonumber\\
&  \left.  +\left(  1+e^{il\pi}\right)  \int d\mathbf{r}d\mathbf{r}^{\prime
}u_{r}\left[  \delta_{\hat{R}_{\phi}\mathbf{r,}\hat{R}_{\phi}\left(
\mathbf{r}^{\prime}-\mathbf{R}_{\delta}\right)  }\left(  \frac{\nabla_{\hat
{R}_{\phi}\mathbf{r}^{\prime}}^{2}}{2m_{e}}+\mu\right)  \right]
e^{il\theta_{\mathbf{r}^{\prime}}}\upsilon_{r^{\prime}}\right\} \nonumber\\
=&\int d\mathbf{r}d\mathbf{r}^{\prime}\left\{  e^{il\phi}\Delta\left(
\mathbf{r-r}^{\prime}+\mathbf{R}_{\delta}\right)  u_{r}u_{r^{\prime}%
}-e^{i2l\phi}e^{il\left(  \theta_{\mathbf{r}}+\theta_{\mathbf{r}^{\prime}%
}\right)  }e^{-il\phi}\Delta^{\ast}\left(  \mathbf{r-r}^{\prime}%
+\mathbf{R}_{\delta}\right)  \upsilon_{r}\upsilon_{r^{\prime}}\right.
\nonumber\\
&  \left.  +\left(  1+e^{il\pi}\right)  \int d\mathbf{r}d\mathbf{r}^{\prime
}u_{r}\left[  \delta_{\mathbf{r,r}^{\prime}+\mathbf{R}_{\delta}}\left(
\frac{\nabla_{\mathbf{r}^{\prime}}^{2}}{2m_{e}}+\mu\right)  \right]
e^{il\phi}e^{il\theta_{\mathbf{r}^{\prime}}}\upsilon_{r^{\prime}}\right\}
\nonumber\\
=& e^{il\phi}\eta\left(  \mathbf{R}_{\delta}\right)  .
\end{align}
Thus the off-diagonal matrix element $t^{+-}$ inherits the rotaional symmetry property of the chiral pairing in the original bulk BdG Hamiltonian.

\begin{figure}
\includegraphics[width=7cm]{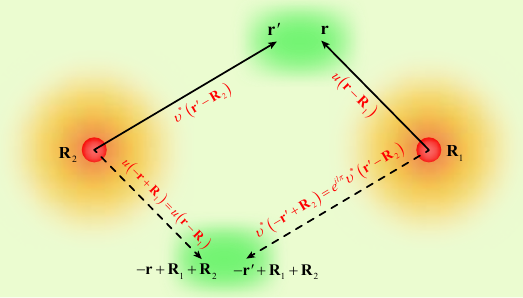}
\caption{Schematic diagram showing the relation between the integrand of Eq.~(\ref{eq:tPMeh}) at two sets of variables: $(\br,\br^\prime)$ indicated by solid arrows and $[\bR_i- (\br^\prime -\bR_j), \bR_j- (\br-\bR_i)]$ in dashed arrows. These two sets are related by a $180^\circ$ rotation about $(\bR_1+\bR_2)/2$.}
\label{fig2}
\end{figure}

Furthermore, in the hybridization between the `$+$' and `$-$' states, the contribution from the electron and hole kinetic processes, $\eta_{ij}^{e}+\eta_{ij}^{h}$,
are sensitive to the parity of the Cooper pairing: $\eta_{ij}^{e}+\eta_{ij}^{h}$ vanishes in odd-parity pairing and is finite in even-parity pairing. This is more obvious in the following expression,
\begin{align}
\eta_{ij}^{e}+\eta_{ij}^{h}=  &  \int d\mathbf{r}d\mathbf{r}^{\prime}\left\{
-u_{\left\vert \mathbf{r-R}_{i}\right\vert }\left[  \delta_{\mathbf{r,r}%
^{\prime}}\left(  -\frac{\nabla_{\mathbf{r}^{\prime}}^{2}}{2m_{e}}-\mu\right)
\right]  e^{il\theta_{\mathbf{r}^{\prime}\mathbf{-R}_{j}}}\upsilon_{\left\vert
\mathbf{r}^{\prime}\mathbf{-R}_{j}\right\vert }+e^{il\theta_{\mathbf{r-R}_{i}%
}}\upsilon_{\left\vert \mathbf{r-R}_{i}\right\vert }\left[  \delta
_{\mathbf{r}^{\prime}\mathbf{,r}}\left(  \frac{\nabla_{\mathbf{r}^{\prime}%
}^{2}}{2m_{e}}+\mu\right)  \right]  u_{\left\vert \mathbf{r}^{\prime
}\mathbf{-R}_{j}\right\vert }\right\} \nonumber\\
=  &  \int d\mathbf{r}d\mathbf{r}^{\prime}\left\{  u_{\left\vert
\mathbf{r-R}_{i}\right\vert }\left[  \delta_{\mathbf{r,r}^{\prime}}\left(
\frac{\nabla_{\mathbf{r}^{\prime}}^{2}}{2m_{e}}+\mu\right)  \right]
e^{il\theta_{\mathbf{r}^{\prime}\mathbf{-R}_{j}}}\upsilon_{\left\vert
\mathbf{r}^{\prime}\mathbf{-R}_{j}\right\vert }+e^{il\theta_{-\mathbf{r}%
^{\prime}\mathbf{+R}_{j}}}\upsilon_{\left\vert -\mathbf{r}^\prime+\mathbf{R}_{j}\right\vert
}\left[  \delta_{\mathbf{r,r}^{\prime}}\left(  \frac{\nabla_{\mathbf{r}}^{2}}{2m_{e}}+\mu\right)  \right]  u_{\left\vert -\mathbf{r+R}%
_{i}\right\vert }\right\} \nonumber\\
=  &  \left(  1+e^{il\pi}\right)  \int d\mathbf{r}d\mathbf{r}^{\prime}%
u_{r}\left[  \delta_{\mathbf{r,r}^{\prime}+\mathbf{R}_{j}-\mathbf{R}_{i}%
}\left(  \frac{\nabla_{\mathbf{r}^{\prime}}^{2}}{2m_{e}}+\mu\right)  \right]
e^{il\theta_{\mathbf{r}^{\prime}}}\upsilon_{r^{\prime}},
\end{align}
which vanishes for odd $l$'s. To obtain the second equation, we made a substitution of variables, $\br \ra  \bR_i- (\br^\prime -\bR_j)$ and $\br^\prime \ra  \bR_j- (\br-\bR_i)$. Pictorially, the two terms in the integrand of the second line are depicted in Fig.~\ref{fig2}. The final expression was obtained after a partial integration and a substitution of variable.

\section{III. Effective tight-binding Hamiltonian \& Anomalous Hall
conductivity}

We are now in position to formally construct the effective tight-binding Hamiltonian for square, triangular and honeycomb superlattices, and study their anomalous Hall effects.

\subsection{A). Square impurity superlattice}
Let us first consider the case with underlying chiral $p$-wave pairing. Following Fig. \ref{fig:TightBinding} (a) and (b) and by Fourier transformation, in square superlattice, the hybridization matrix in the momentum space can be expressed as $\hat{h}_{\mathbf{k}}=\sum_{\delta}e^{i\mathbf{k\cdot\mathbf{R}%
}_{\delta}}\hat{h}\left(  \mathbf{R}_{\delta}\right)  $, in which the matrix elements with only considering the nearest-neighbor terms are expressed as
\begin{equation}
\lambda_{\mathbf{k}}=2\lambda\left(  \cos k_{x}+\cos k_{y}\right),~
\eta_{\mathbf{k}}=2\eta\left(  \sin k_{x}+i\sin k_{y}\right),%
\end{equation}
where $\lambda\equiv \lambda(R_{0})$, $\eta\equiv \eta(R_{0}\mathbf{\hat{y}})$ are real constants. The decomposition $\lambda = \lambda^\Delta + \lambda^e + \lambda^h$ and $\eta = \eta^\Delta + \eta^e + \eta^h$ are implicit. Note that, for brevity, we have suppressed $R_{0}$ in $\lambda_{\mathbf{k}}$, $\eta_{\mathbf{k}}$ and hereafter. Then, the effective Hamiltonian for the impurity superlattice follows as,
\begin{equation}
H_{\mathbf{k}}^{\text{eff}}=\mathcal{E}_{3\mathbf{k}}\sigma_{3}+\mathcal{E}%
_{1\mathbf{k}}\sigma_{1}-\mathcal{E}_{2\mathbf{k}}\sigma_{2},
\label{eq:suppHeff}
\end{equation}
in which%

\begin{equation}
\mathcal{E}_{3\mathbf{k}} =E_{0}+2\lambda\left(  \cos k_{x}+\cos
k_{y}\right),~~\mathcal{E}_{1\mathbf{k}}=2\eta\sin k_{x},~~
\mathcal{E}_{2\mathbf{k}} =2\eta\sin k_{y}.
\end{equation}
This effective Hamiltonian resembles the original chiral p-wave model.
As a side remark, the idea to design topological band structure through super-modulations of the order parameter is not new. Besides the present model which also appeared in Ref.~\cite{Kimme16}, a superlattice of magnetic impurities in a conventional superconductor with Rashiba spin-orbit coupling has also been shown to support subgap bands with high Chern numbers~\cite{Rontynen15}. In another context, a pair-density-wave of a chiral $p$-wave order parameter was shown to generate topologically protected low-energy excitations~\cite{Santos19}.

Similarly, following the argument given in the main text as well as Fig. \ref{fig:TightBinding} (c) and (d), the matrix elements for effective current operators along the $x$- and $y$-directions take the following forms:
\begin{align}
J_{x\mathbf{k}}^{++}=&-2\left(\lambda^{e}-\lambda^{h}\right)  \sin k_{x},~J_{x\mathbf{k}}^{+-}=2(\eta^{e}-\eta^{h})\cos k_{x}=4\eta^{e}\cos k_{x},\\
J_{y\mathbf{k}}^{++}=&-2\left(\lambda^{e}-\lambda^{h}\right) \sin k_{y},~
J_{y\mathbf{k}}^{+-}=2(\eta^{e}-\eta^{h})\cos k_{y}=4\eta^{e}\cos k_{y},
\end{align}
Hence the $i$-th component of the current operators in terms of the Pauli matrices can be written as:
\begin{equation}
\hat{J}_{i\mathbf{k}}^{\text{eff}}=\mathcal{J}_{3i\mathbf{k}}\sigma_{3}+\mathcal{J}_{1i\mathbf{k}}
\sigma_{1}+\mathcal{J}_{2i\mathbf{k}} \sigma_{2},
\end{equation}
in which%
\begin{equation}
\mathcal{J}_{3x\mathbf{k}}=-2\left(  \lambda^{e}-\lambda^{h}\right)  \sin k_{x},~
\mathcal{J}_{1x\mathbf{k}}=4\eta^{e}\cos k_{x},~
\mathcal{J}_{2x\mathbf{k}}=0,
\end{equation}
for $x$-direction, and
\begin{equation}
\mathcal{J}_{3y\mathbf{k}}=-2\left(  \lambda^{e}-\lambda^{h}\right)\sin k_{y},~
\mathcal{J}_{1y\mathbf{k}}=0,~
\mathcal{J}_{2y\mathbf{k}}=-4\eta^{e}\cos k_{y},
\end{equation}
for $y$-direction. Within linear-response theory, the transverse current-current
correlation function at one-loop level is given by,
\begin{equation}
\pi_{xy}\left(  \mathbf{q},i\nu_{m}\right)  =T\sum_{\mathbf{k},i\omega_{n}%
}\text{Tr}\left[  \hat{J}_{x\mathbf{k}}^{\text{eff}}  \hat
{G}\left(  \mathbf{k+q},i\omega_{n}+i\nu_{m}\right)  \hat{J}_{y\mathbf{k}}^{\text{eff}%
} \hat{G}\left(  \mathbf{k},i\omega_{n}\right)
\right]  ,
\end{equation}
where $T$ is the temperature, $\omega_{n}=\left(  2n+1\right)  \pi T$ and $\nu_{m}=2m\pi T$ are the fermionic and bosonic Matsubara frequencies, respectively. $\hat{G}\left(  \mathbf{k},i\omega_{n}\right)  $ is the single-particle Green's function which can be written as
\begin{equation}
\hat{G}\left(  \mathbf{k},i\omega_{n}\right)  =\left(  i\omega_{n}\sigma
_{0}-H_{\mathbf{k}}^{\text{eff}}\right)  ^{-1}=\frac{i\omega_{n}\sigma
_{0}+\mathcal{E}_{3\mathbf{k}}\sigma_{3}+\mathcal{E}_{1\mathbf{k}}\sigma
_{1}+\mathcal{E}_{2\mathbf{k}}\sigma_{2}}{\left(  i\omega_{n}\right)
^{2}-E_{\mathbf{k}}^{2}},
\end{equation}
where $E_{\mathbf{k}}=\sqrt{\mathcal{E}^2_{3\mathbf{k}}+\mathcal{E}^2_{1\mathbf{k}} +\mathcal{E}^2_{2\mathbf{k}}}$ is the quasiparticle dispersion.

The Hall conductivity is given by the antisymmetric part of the transverse current correlator. After some algebra and an analytical continuation $i\nu_{m}\rightarrow\omega+i\delta$, we arrive at the following, %
\begin{equation}
\sigma_{\text{H}}\left(  \omega+i\delta\right)  =\frac{i}{2\omega}%
\lim_{\mathbf{q}\rightarrow0}\left[  \pi_{xy}\left(  \mathbf{q},\omega
+i\delta\right)  -\pi_{yx}\left(  \mathbf{q},\omega+i\delta\right)  \right]
=\sum_{\mathbf{k}}\frac{f\left(  \mathbf{k}\right)  }{E_{\mathbf{k}}\left[
\left(  \omega+i\delta\right)  ^{2}-4E_{\mathbf{k}}^{2}\right]  },
\end{equation}
in which
\begin{equation}
f\left(  \mathbf{k}\right)  =
\sum_{s,m,n}\frac{\epsilon^{smn}}{2}\left[  \mathcal{J}_{sx\mathbf{k}} \mathcal{J}_{my\mathbf{k}} -\mathcal{J}%
_{sy\mathbf{k}} \mathcal{J}_{my\mathbf{k}}
\right]  \mathcal{E}_{n\mathbf{k}}.
\end{equation}
Substituting the expressions, we see that a non-zero anomalous Hall conductivity emerges in the impurity superlattice embedded in a chiral p-wave superconductor.

\begin{figure}
\includegraphics[width=7cm]{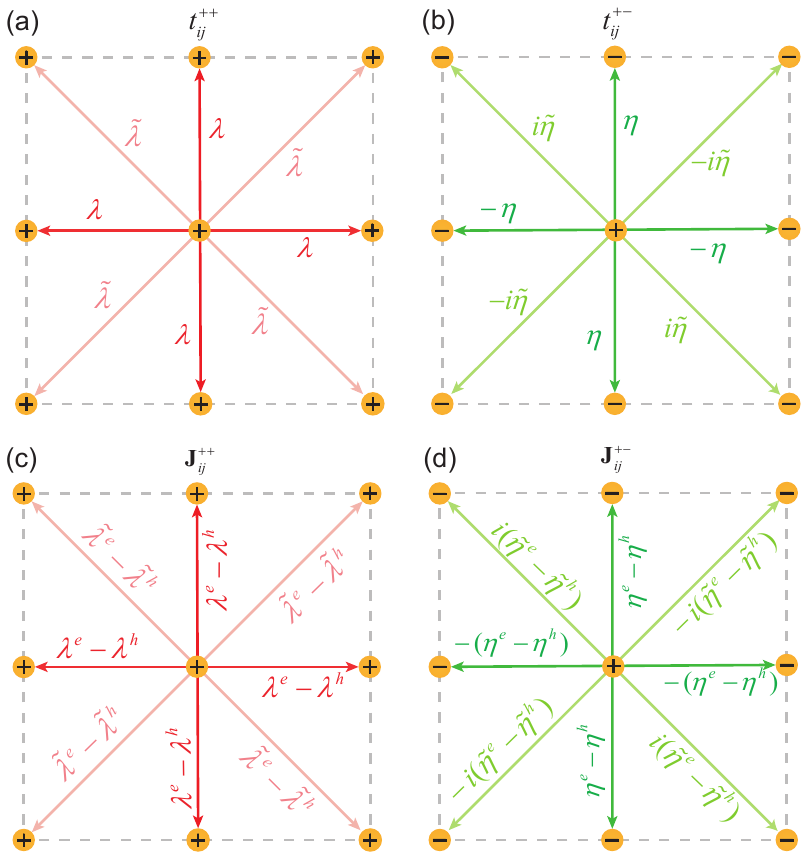}
\caption{ (a) (b) Tight-binding construction of a square impurity superlattice embedded in a chiral d-wave superconductor. Note the relation $\lambda= \lambda^e+\lambda^h+\lambda^\Delta$ and the same for $\tilde{\lambda}$, $\eta$ and $\tilde{\eta}$. (c) (d) The current operator on the superlattice. The symbols `$+$' and `$-$' symbols designate the impurity bound states, and arrows indicate the reference direction of hopping or current flow.}
\label{fig:TightBindingD}
\end{figure}

We now turn to the case of underlying chiral $d$-wave pairing. We find that, a full description of low-energy model requires a consideration of up to the next-nearest neighboring terms shown in Fig.~\ref{fig:TightBindingD}, after which we obtain,
\begin{equation}
\lambda_{\mathbf{k}}=2\lambda\left(  \cos k_{x}+\cos k_{y}\right)  +4\tilde{\lambda}\cos k_{x}\cos k_{y},~~\eta_{\mathbf{k}}=-2\eta\left(  \cos k_{x}-\cos k_{y}\right)  -i4\tilde{\eta}\sin k_{x}\sin k_{y},
\end{equation}
in which $\tilde{\lambda}\equiv \lambda(\sqrt{2}R_{0})$ and $\tilde{\eta}\equiv \eta(\sqrt{2}R_{0}\mathbf{\hat{y}})$ are hopping integrals associated with the next-nearest neighboring contributions. Written in the form of Eq.~(\ref{eq:suppHeff}), the corresponding $\mathcal{E}_{i\mathbf{k}}$ are given by
\begin{equation}
\mathcal{E}_{3\mathbf{k}} =E_{0}+2\lambda\left(  \cos k_{x}+\cos
k_{y}\right)+4\tilde{\lambda}\cos k_x \cos k_y,~~\mathcal{E}_{1\mathbf{k}}=-2\eta (\cos k_{x}-\cos k_y ),~~\mathcal{E}_{2\mathbf{k}} =4\tilde{\eta}\sin k_x\sin k_{y}.
\end{equation}
Turning to the current operators, we have $\hat{J}^{+-}_{x\mathbf{k}}= \hat{J}^{+-}_{y\mathbf{k}}=0$ on account of the parity constraints ($\eta^e-\eta^h=0$ for underlying even-parity pairing) discussed in the previous section. Thus $\hat{J}_{i\mathbf{k}}^{\text{eff}}=\mathcal{J}_{3i\mathbf{k}}\sigma_{3}=-2\lambda^e\cos k_i\sigma_{3}$ with $i=x,y$. A straightforward calculation shows that the resultant model generate no anomalous Hall conductivity at the one-loop calculation.

\subsection{B). Triangular impurity superlattice}
In the case of triangular impurity superlattices, the anomalous Hall conductivity has the same form as in the case of a square superlattice, but with slight modifications. Consider only the nearest-neighbor hoppings, one obtains,
\begin{subequations}
\begin{align}
\mathcal{E}_{3\mathbf{k}} &  =E_{0}+2\lambda\left(  \cos k_{x}+2\cos
\frac{k_{x}}{2}\cos\frac{\sqrt{3}k_{y}}{2}\right)  ,\\
\mathcal{E}_{1\mathbf{k}} &  =\left\{
\begin{array}
[c]{cc}%
2\eta\left(  \sin k_{x}+\sin\frac{k_{x}}{2}\cos\frac{\sqrt{3}k_{y}}%
{2}\right)  , & \text{ }\left(  l=1\right)  \\
2\eta\left(  \sin k_{x}-\cos\frac{k_{x}}{2}\cos\frac{\sqrt{3}k_{y}}%
{2}\right)  , & \text{ }\left(  l=2\right)
\end{array}
\right.  \\
\mathcal{E}_{2\mathbf{k}} &  =\left\{
\begin{array}
[c]{cc}%
2\sqrt{3}\eta\cos\frac{k_{x}}{2}\sin\frac{\sqrt{3}k_{y}}{2}, & \text{
}\left(  l=1\right)  \\
2\sqrt{3}\eta\sin\frac{k_{x}}{2}\sin\frac{\sqrt{3}k_{y}}{2}, & \text{
}\left(  l=2\right)
\end{array}
\right.  ,
\end{align}
and the associated components of the current operators are,%
\end{subequations}
\begin{subequations}
\begin{align}
\mathcal{J}_{3x\mathbf{k}} &  =-2\left(  \lambda^{e}-\lambda^{h}\right)  \left(  \sin k_{x}+\sin\frac{k_{x}}{2}\cos
\frac{\sqrt{3}k_{y}}{2}\right)  ,\\
\mathcal{J}_{1x\mathbf{k}}  &  =\left\{
\begin{array}
[c]{cc}%
4\eta^{e}\left(  \cos k_{x}+\frac{1}{2}%
\cos\frac{k_{x}}{2}\cos\frac{\sqrt{3}k_{y}}{2}\right)  , & \text{ }\left(
l=1\right)  \\
0 , & \text{ }\left(
l=2\right)
\end{array}
\right.  \\
\mathcal{J}_{2x\mathbf{k}}  &  =\left\{
\begin{array}
[c]{cc}%
-6\eta^{e}\sin\frac{k_{x}%
}{2}\cos\frac{\sqrt{3}k_{y}}{2}, & \text{ }\left(  l=1\right)  \\
0, & \text{ }\left(  l=2\right)
\end{array}
\right.
\end{align}
and%
\end{subequations}
\begin{subequations}
\begin{align}
\mathcal{J}_{3y\mathbf{k}}   &  =-2\sqrt{3}\left(  \lambda^{e}-\lambda^{h}\right)  \cos\frac{k_{x}}{2}\sin\frac{\sqrt{3}k_{y}%
}{2},\\
\mathcal{J}_{1y\mathbf{k}}  &  =\left\{
\begin{array}
[c]{cc}%
-6\eta^{e}\sin\frac{k_{x}}{2}\sin\frac{\sqrt{3}k_{y}}{2}, & \text{ }\left(  l=1\right)  \\
0, & \text{ }\left(  l=2\right)
\end{array}
\right.  \\
\mathcal{J}_{2y\mathbf{k}} &  =\left\{
\begin{array}
[c]{cc}%
6\eta^{e}\cos\frac{k_{x}}{2}%
\cos\frac{\sqrt{3}k_{y}}{2}, & \text{ }\left(  l=1\right)  \\
0, & \text{ }\left(  l=2\right)
\end{array}
\right.
\end{align}
It can thus be seen that the both the effective tight-binding Hamiltonian and the effective current operators follows the same overall structure as those in the square superlattice models. One thus expects the same outcome as far as the anomalous Hall effect is concerned.

\subsection{C). Honeycomb impurity superlattice}
The honeycomb impurity superlattice is very different from the previous two cases, since the enlargement of the Hilbert space due to an added sublattice degree of freedom. Consider a basis $\Psi_{i}=\left(  c_{i,+},c_{i,+}^{\prime},c_{i,-},c_{i,-}^{\prime}\right)  ^{\intercal}$, with $c$ and $c^{\prime}$ representing the two sublattices, a general effective Hamiltonian for impurity superlattice with nearest-neighbor hoppings has the following form
\end{subequations}
\begin{equation}
\hat{H}^{\text{eff}}_{\bk}=
\begin{bmatrix}
    E_0 & \lambda_{\bk} & 0 & \eta_{\bk}\\
    \lambda_{\bk}^{*} & E_0 & (-1)^l\eta_{-\bk} & 0\\
    0 & (-1)^l\eta_{-\bk}^{*}& -E_0 &-\lambda_{\bk} \\
    \eta_{\bk}^{*} & 0 &-\lambda_{\bk}^{*} & -E_0
\end{bmatrix}
\end{equation}
where the matrix elements are given by,%
\begin{align}
\lambda_{\mathbf{k}}  &  =\sum_{\delta}e^{i\mathbf{k\cdot R}_{\delta}}%
\lambda=\lambda\left(  1+2e^{-i\frac{3k_{x}}{2}}\cos\frac{\sqrt
{3}k_{y}}{2}\right)  ,\\
\eta_{\mathbf{k}}  &  =\sum_{\delta}e^{i\mathbf{k\cdot R}_{\delta}}%
e^{il\theta_{\mathbf{R}_{\delta}}}\eta \xrightarrow{l=2}~\eta\left[  1+2e^{-i\frac
{3k_{x}}{2}}\cos\left(  \frac{\sqrt{3}k_{y}}{2}-\frac{2\pi}{3}\right)
\right]  .
\end{align}
in which we have eliminated a prefactor $e^{ik_{x}}$ by a standard gauge transformation similar to the treatment for monolayer graphene. Note also that in the last expression, we have explicitly taken the example of $l=2$ for underlying chiral d-wave, and the same below. The current operator follows as,%
\begin{align}
\hat{J}_{x\mathbf{k}}^{\text{eff}}  &  =\left(
\begin{array}
[c]{cccc}%
0 & J_{x\mathbf{k}}^{++}  & 0 & J_{x\mathbf{k}}^{+-} \\
\left(  J_{x\mathbf{k}}^{++} \right)  ^{\ast} & 0 &
-J_{x,\mathbf{-k}}^{+-} & 0\\
0 & -\left(  J_{x,\mathbf{-k}}^{+-} \right)  ^{\ast} & 0 &
-J_{x\mathbf{k}}^{++} \\
\left(  J_{x\mathbf{k}}^{+-} \right)  ^{\ast} & 0 & -\left(
J_{x\mathbf{k}}^{++} \right)  ^{\ast} & 0
\end{array}
\right) \nonumber\\
&  =\mathcal{J}_{1x\mathbf{k}} \varrho_{1}\otimes\sigma
_{3}+\mathcal{J}_{2x\mathbf{k}} \varrho_{2}\otimes\sigma
_{3}+\mathcal{J}_{3x\mathbf{k}} \varrho_{1}\otimes\sigma
_{1}+\mathcal{J}_{4x\mathbf{k}} \varrho_{1}\otimes\sigma
_{2}+\mathcal{J}_{5x\mathbf{k}}  \varrho_{2}\otimes\sigma
_{1}+\mathcal{J}_{6x\mathbf{k}}  \varrho_{2}\otimes\sigma_{2},
\end{align}
or, equivalently,
\begin{equation}
\hat{J}_{y\mathbf{k}}^{\text{eff}}  =\mathcal{J}_{1y\mathbf{k}} \varrho_{1}\otimes\sigma_{3}+\mathcal{J}_{2y\mathbf{k}}  \varrho_{2}\otimes\sigma_{3}+\mathcal{J}_{3y\mathbf{k}} \varrho_{1}\otimes\sigma_{1}+\mathcal{J}_{4y\mathbf{k}} \varrho_{1}\otimes\sigma_{2}+\mathcal{J}_{5y\mathbf{k}} \varrho_{2}\otimes\sigma_{1}+\mathcal{J}_{6y\mathbf{k}} \varrho_{2}\otimes\sigma_{2},
\end{equation}
where
\begin{align}
J_{x\mathbf{k}}^{++}  &  =\sum_{\delta}e^{i\mathbf{k\cdot
\mathbf{R}}_{\delta}}J_{x}^{++}\left(  \mathbf{R}_{\delta}\right)  =-3\left(
\lambda^{e}-\lambda^{h}\right)  \left(  \sin\frac{3k_{x}}{2}\cos
\frac{\sqrt{3}k_{y}}{2}+i\cos\frac{3k_{x}}{2}\cos\frac{\sqrt{3}k_{y}}%
{2}\right)  ,\\
J_{y\mathbf{k}}^{++} &  =\sum_{\delta}e^{i\mathbf{k\cdot
\mathbf{R}}_{\delta}}J_{y}^{++}\left(  \mathbf{R}_{\delta}\right)  =-\sqrt
{3}\left(  \lambda^{e}-\lambda^{h}\right)  \left(  \cos\frac{3k_{x}%
}{2}\sin\frac{\sqrt{3}k_{y}}{2}-i\sin\frac{3k_{x}}{2}\sin\frac{\sqrt{3}k_{y}%
}{2}\right)  ,\\
J_{x\mathbf{k}}^{+-}  &  =\sum_{\delta}e^{i\mathbf{k\cdot
\mathbf{R}}_{\delta}}J_{x}^{+-}\left(  \mathbf{R}_{\delta}\right)  =\left\{
\begin{array}
[c]{cc}%
-3\left(  \eta^{e}-\eta^{h}\right)  \left(  \sin\frac{3k_{x}}{2}%
+i\cos\frac{3k_{x}}{2}\right)  \cos\left(  \frac{\sqrt{3}k_{y}}{2}-\frac{2\pi
}{3}\right)  , & \text{ }\left(  l=1\right) \\
0, & \text{ }\left(  l=2\right)
\end{array}
\right. \\
J_{y\mathbf{k}}^{+-}  &  =\sum_{\delta}e^{i\mathbf{k\cdot
\mathbf{R}}_{\delta}}J_{y}^{++}\left(  \mathbf{R}_{\delta}\right)  =\left\{
\begin{array}
[c]{cc}%
-\sqrt{3}\left(  \eta^{e}-\eta^{h}\right)  \left(  \cos\frac{3k_{x}%
}{2}-i\sin\frac{3k_{x}}{2}\right)  \sin\left(  \frac{\sqrt{3}k_{y}}{2}%
-\frac{2\pi}{3}\right)  , & \text{ }\left(  l=1\right) \\
0, & \text{ }\left(  l=2\right)
\end{array}
\right.
\end{align}
and $\mathcal{J}_{1i\mathbf{k}} =\operatorname{Re}(
J_{i\mathbf{k}}^{++})  $, $\mathcal{J}_{2i\mathbf{k}} =-\operatorname{Im}(  J_{i\mathbf{k}}^{++})  $, $\mathcal{J}_{3i\mathbf{k}}
=\operatorname{Re}( J_{i\mathbf{k}}^{+-} -J_{i,-\mathbf{k}}^{+-})/2$, $\mathcal{J}_{4i\mathbf{k}} =-\operatorname{Im}( J_{i\mathbf{k}}^{+-} -J_{i,-\mathbf{k}}^{+-}) /2$, and $\mathcal{J}_{5i\mathbf{k}} =-\operatorname{Im}(J_{i\mathbf{k}}^{+-} +J_{i\mathbf{k}}^{+-})/2$, $\mathcal{J}_{6i\mathbf{k}}=-\operatorname{Re}( J_{i\mathbf{k}}^{+-}+J_{i,-\mathbf{k}}^{+-})/2$, with $i=x,y$.

The Green's function $\hat{G}\left(  \mathbf{k},i\omega_{n}\right)  =\left(  i\omega_{n}\sigma_{0}-H_{\mathbf{k}}^{\text{eff}}\right)  ^{-1}$ acquires the following form,
\begin{equation}
\hat{G}\left(  \mathbf{k},i\omega_{n}\right)  =\sum_{i,j=0,1,2,3}\frac
{g_{ij}\varrho_{i}\otimes\sigma_{j}}{\left[  \left(  i\omega_{n}\right)
^{2}-E_{+,\mathbf{k}}^{2}\right]  \left[  \left(  i\omega_{n}\right)
^{2}-E_{-,\mathbf{k}}^{2}\right]  },
\end{equation}
where $g_{00}=-i\omega_{n}(\omega_{n}^{2}+E_{0}^{2}+\left\vert \lambda
_{\mathbf{k}}\right\vert ^{2}+\frac{\left\vert \eta_{\mathbf{k}}\right\vert
^{2}+\left\vert \eta_{-\mathbf{k}}\right\vert ^{2}}{2})$, $g_{03}%
=-E_{0}(\omega_{n}^{2}+E_{0}^{2}-\left\vert \lambda_{\mathbf{k}}\right\vert
^{2}+\frac{\left\vert \eta_{\mathbf{k}}\right\vert ^{2}+\left\vert
\eta_{-\mathbf{k}}\right\vert ^{2}}{2})$, $g_{33}=-i\omega_{n}(\frac
{\left\vert \eta_{\mathbf{k}}\right\vert ^{2}-\left\vert \eta_{-\mathbf{k}%
}\right\vert ^{2}}{2})$, $g_{30}=-E_{0}(\frac{\left\vert \eta_{\mathbf{k}%
}\right\vert ^{2}-\left\vert \eta_{-\mathbf{k}}\right\vert ^{2}}{2})$,
$g_{11}=-\frac{1}{2}\operatorname{Re}[\eta_{-\mathbf{k}}(\omega_{n}^{2}%
+E_{0}^{2}+\left\vert \eta_{\mathbf{k}}\right\vert ^{2}+\lambda_{\mathbf{k}%
}^{2})+\eta_{\mathbf{k}}(\omega_{n}^{2}+E_{0}^{2}+\left\vert \eta
_{-\mathbf{k}}\right\vert ^{2}+\lambda_{\mathbf{k}}^{\ast2})]$, $g_{12}%
=-\frac{1}{2}\operatorname{Im}[\eta_{-\mathbf{k}}(\omega_{n}^{2}+E_{0}%
^{2}+\left\vert \eta_{\mathbf{k}}\right\vert ^{2}+\lambda_{\mathbf{k}}%
^{2})+\eta_{\mathbf{k}}(\omega_{n}^{2}+E_{0}^{2}+\left\vert \eta_{-\mathbf{k}%
}\right\vert ^{2}+\lambda_{\mathbf{k}}^{\ast2})]$, $g_{21}=\frac{1}%
{2}\operatorname{Im}[\eta_{-\mathbf{k}}(\omega_{n}^{2}+E_{0}^{2}-\left\vert
\eta_{\mathbf{k}}\right\vert ^{2}-\lambda_{\mathbf{k}}^{2})+\eta_{\mathbf{k}%
}(\omega_{n}^{2}+E_{0}^{2}+\left\vert \eta_{-\mathbf{k}}\right\vert
^{2}-\lambda_{\mathbf{k}}^{\ast2})]$, $g_{22}=-\frac{1}{2}\operatorname{Re}%
[\eta_{-\mathbf{k}}(\omega_{n}^{2}+E_{0}^{2}+\left\vert \eta_{\mathbf{k}%
}\right\vert ^{2}-\lambda_{\mathbf{k}}^{2})-\eta_{\mathbf{k}}(\omega_{n}%
^{2}+E_{0}^{2}+\left\vert \eta_{-\mathbf{k}}\right\vert ^{2}-\lambda
_{\mathbf{k}}^{\ast2})]$, $g_{13}=-\operatorname{Re}[\lambda_{\mathbf{k}%
}(\omega_{n}^{2}-E_{0}^{2}+\left\vert \lambda_{\mathbf{k}}\right\vert
^{2}-\eta_{\mathbf{k}}^{\ast}\eta_{-\mathbf{k}})]$, $g_{23}=-\operatorname{Im}%
[\lambda_{\mathbf{k}}(\omega_{n}^{2}-E_{0}^{2}+\left\vert \lambda_{\mathbf{k}%
}\right\vert ^{2}-\eta_{\mathbf{k}}^{\ast}\eta_{-\mathbf{k}})]$,
$g_{10}=2i\omega_{n}E_{0}\operatorname{Re}\left(  \lambda_{\mathbf{k}}\right)
$, $g_{20}=2i\omega_{n}E_{0}\operatorname{Im}\left(  \lambda_{\mathbf{k}%
}\right)  $, $g_{31}=-i\omega_{n}\operatorname{Re}\left(  \lambda_{\mathbf{k}%
}\eta_{\mathbf{k}}^{\ast}-\lambda_{\mathbf{k}}^{\ast}\eta_{-\mathbf{k}}^{\ast
}\right)  $, $g_{32}=i\omega_{n}\operatorname{Im}\left(  \lambda_{\mathbf{k}%
}\eta_{\mathbf{k}}^{\ast}-\lambda_{\mathbf{k}}^{\ast}\eta_{-\mathbf{k}}^{\ast
}\right)  $, $g_{01}=E_{0}\operatorname{Re}\left(  \lambda_{\mathbf{k}}%
\eta_{\mathbf{k}}^{\ast}+\lambda_{\mathbf{k}}^{\ast}\eta_{-\mathbf{k}}^{\ast
}\right)  $, $g_{32}=-E_{0}\operatorname{Im}\left(  \lambda_{\mathbf{k}}%
\eta_{\mathbf{k}}^{\ast}+\lambda_{\mathbf{k}}^{\ast}\eta_{-\mathbf{k}}^{\ast
}\right)  $, and $E_{\pm,\mathbf{k}}=\sqrt{E_{0}^{2}+\left\vert \lambda_{\mathbf{k}}\right\vert
^{2}+\frac{1}{2}\left(  \left\vert \eta_{\mathbf{k}}\right\vert ^{2}%
+\left\vert \eta_{-\mathbf{k}}\right\vert ^{2}\right)  \pm\sqrt{4\left\vert
\lambda_{\mathbf{k}}\right\vert ^{2}E_{0}^{2}+\left\vert \lambda_{\mathbf{k}}^{\ast}\eta_{\mathbf{k}}-\lambda_{\mathbf{k}%
}\eta_{-\mathbf{k}}\right\vert
^{2}+\frac{1}{4}\left(  \left\vert \eta_{\mathbf{k}}\right\vert ^{2}%
-\left\vert \eta_{-\mathbf{k}}\right\vert ^{2}\right)  ^{2}}}$.

We mainly focus on the case with chiral $d$-wave (even-parity) pairing in which $\mathcal{J}_{3i}=\mathcal{J}%
_{4i}=\mathcal{J}_{5i}=\mathcal{J}_{6i}=0$ ($i=x,y$), and study its anomalous Hall conductivity. A lengthy calculation leads to,
\begin{align}
\pi &  _{xy}\left(  \mathbf{q=0},i\nu_{m}\right)  -\pi_{yx}\left(
\mathbf{q=0},i\nu_{m}\right) \nonumber\\
= &  \sum_{\mathbf{k}}\frac{\nu_{m}}{E_{+}E_{-}}\left\{  \frac{f\left(
E_{+}\right)  -f\left(  E_{-}\right)  }{\left(  E_{+,\mathbf{k}}%
-E_{-,\mathbf{k}}\right)  \left[  \left(  E_{+,\mathbf{k}}-E_{-,\mathbf{k}%
}\right)  ^{2}+\nu_{m}^{2}\right]  }+\frac{1-f\left(  E_{+}\right)  -f\left(
E_{-}\right)  }{\left(  E_{+,\mathbf{k}}+E_{-,\mathbf{k}}\right)  \left[
\left(  E_{+,\mathbf{k}}+E_{-,\mathbf{k}}\right)  ^{2}+\nu_{m}^{2}\right]
}\right\} \nonumber\\
&  \times E_{0}\left(  \left\vert \eta_{\mathbf{k}}\right\vert ^{2}-\left\vert
\eta_{-\mathbf{k}}\right\vert ^{2}\right)  \left[  \mathcal{J}_{1x}\left(
\mathbf{k}\right)  \mathcal{J}_{2y}\left(  \mathbf{k}\right)  -\mathcal{J}%
_{2x}\left(  \mathbf{k}\right)  \mathcal{J}_{1y}\left(  \mathbf{k}\right)
\right]  .
\end{align}
It returns the following zero-temperature anomalous Hall conductivity at real frequency:
\begin{equation}
\sigma_{\text{H}}\left(  \omega+i\delta\right)  =\sum_{\mathbf{k}}\frac
{E_{0}\left(  \left\vert \eta_{\mathbf{k}}\right\vert ^{2}-\left\vert
\eta_{-\mathbf{k}}\right\vert ^{2}\right)  \left[  \mathcal{J}_{1x}\left(
\mathbf{k}\right)  \mathcal{J}_{2y}\left(  \mathbf{k}\right)  -\mathcal{J}%
_{2x}\left(  \mathbf{k}\right)  \mathcal{J}_{1y}\left(  \mathbf{k}\right)
\right]  }{2E_{+,\mathbf{k}}E_{-,\mathbf{k}}\left(  E_{+,\mathbf{k}%
}+E_{-,\mathbf{k}}\right)  \left[  \left(  E_{+,\mathbf{k}}+E_{-,\mathbf{k}%
}\right)  ^{2}-\left(  \omega+i\delta\right)  ^{2}\right]  }.
\end{equation}
This quantity is generically finite. Hence, distinct from cases of square and triangular superlattices, the anomalous Hall conductivity for chiral $d$-wave and other even-parity chiral states is finite. One can further check that the honeycomb superlattice models with underlying odd-parity chiral pairings also support finite Hall conductance.


\begin{thebibliography}{99}
\bibitem{footnote2} Notice that the form of this hybridization is locked with the $U(1)$ phase of the underlying chiral pairing. For example, for chiral pairings with $\Delta_{\bk} \sim (k_x+ik_y)^l$ and $\Delta_{\bk} \sim e^{i\varphi}(k_x+ik_y)^l$, their $\eta_{ij}$ must differ by the same phase $\varphi$.
\end{thebibliography}
\end{document}